\documentclass[aps, prb, 10pt, superscriptaddress, twocolumn]{revtex4-1}
	\usepackage[bookmarks=false,linkcolor=blue,urlcolor=blue,colorlinks,citecolor=blue]{hyperref}

	\usepackage{graphicx}
	\usepackage{color}
	\usepackage{amsmath}
	\usepackage[normalem]{ulem}
	\usepackage[toc,page]{appendix}
	\usepackage{amsfonts}
	\usepackage{amssymb}

\newcommand{\be}{\begin{equation}}
\newcommand{\ee}{\end{equation}}
\newcommand{\p}{\partial}

\newcommand{\la}{\label}
\newcommand{\bea}{\begin{eqnarray}}
\newcommand{\eea}{\end{eqnarray}}

\begin{document}

\title{\begin{flushright}\vspace{-1in}
			\mbox{\normalsize  }
		\end{flushright}
	Towards classification of Fracton phases: the multipole algebra  
	 \vskip 20pt
	 }

\author{Andrey Gromov}
\affiliation{
UC Berkeley \& Lawrence Berkeley National Laboratory, Berkeley, California 94720, USA}

\date{\today}

\begin{abstract}

We present an effective field theory approach to the Fracton phases. The approach is based the notion of a multipole algebra. It is an extension of space(-time) symmetries of a charge-conserving matter that includes global symmetries responsible for the conservation of various components of the multipole moments of the charge density. We explain how to construct field theories invariant under the action of the algebra. These field theories generally break rotational invariance and exhibit anisotropic scaling. We further explain how to partially gauge the multipole algebra. Such gauging makes the symmetries responsible for the conservation of multipole moments local, while keeping rotation and translations symmetries global. It is shown that upon such gauging one finds the symmetric tensor gauge theories, as well as the generalized gauge theories discussed recently in the literature. The outcome of the gauging procedure depends on the choice of the multipole algebra. In particular, we show how to construct an effective theory for the $U(1)$ version of the Haah code based on the principles of symmetry and provide a two dimensional example with operators supported on a Sierpinski triangle. We  show that upon condensation of charged excitations Fracton phases of both types as well as various SPTs emerge. Finally, the relation between the present approach and the formalism based on polynomials over finite fields is discussed. 

\end{abstract}


\maketitle


\section{Introduction} 

Fracton order is a class of gapped phases of matter that exhibits a system-size dependent groundstate degeneracy on a space of non-trivial topology. This degeneracy cannot be lifted by local perturbations. Another striking feature of the fracton order is the existence of local topologically non-trivial\footnote{That cannot be created by a local operator} excitations  with ``restricted mobility''\,\cite{Chamon2005,haah2011local, bravyi2011topological, yoshida2013exotic, vijay2015new, vijay2016fracton}. The latter refers to the absence of string-like operators that would allow the local excitations to move through the space without creating additional excitations. This is in sharp contrast with, for example, fractional quantum Hall states, where quasiholes can be freely transported (provided they were localized by an external potential) \footnote{System size-dependent degeneracy has appeared in quantum Hall context previously \cite{kol1993fractional}}. These models were originally introduced  as an example of glassy behavior without disorder by Chamon and as a model for stable quantum memory by Haah. The term ``\emph{fracton}'' has been previously used to refer to small scale thermal vibrations of fractal structures \cite{alexander1982density}. We hope that the present use of the term will not cause confusion. 

According to the recent nomenclature\cite{vijay2016fracton}, the fracton phases come in two varieties: type-I and type-II. Type-I phases support completely immobile excitations -- fractons -- at corners of codimension $1$ surface operators (such as membranes in $d=3$) \cite{Chamon2005, yoshida2013exotic, vijay2015new, vijay2016fracton}. Various combinations of fractons can freely move around on lower dimensional sub-manifolds \cite{vijay2015new, vijay2016fracton}. Type-II phases support only immobile fracton excitations, that exist at ``corners'' of fractal operators, \emph{i.e.} the non-local operators, supported on a fractal.

Despite significant research effort it is presently not known what is the appropriate mathematical structure that encodes the exotic properties of the fracton order in a model-independent fashion. First substantial progress in model-independent description of fractons was made in a series of papers [\onlinecite{pretko2017generalized,pretko2017subdimensional, pretko2017emergent}], where it was explained that restricted mobility of type-I models can be incorporated into an effective field theory by enforcing a certain set of Gauss law constraints. These constraints lead to the conservation of the dipole moment (or, generally various multipole moments) of the matter fields. It was also shown that lattice gauge theories with such Gauss law constraints have been previously studied in [\onlinecite{xu2006novel}-\onlinecite{xu2010emergent}]. Degrees of freedom in these theories are described by a \emph{symmetric tensor} gauge field and the Gauss law is enforced upon a symmetric tensor electric field. This type of effective theories does not describe a gapped phase ``as is'', since there are gapless excitations. A version of Higgs mechanism was developed to remove the gapless modes in [\onlinecite{BB2018Higgs}-\onlinecite{ma2018fracton}]. 

In a somewhat surprising parallel development it was argued that certain type-I fracton models are related to a quantum theory of elasticity \cite{pretko2018fracton, gromov2017fractional, kumar2018symmetry, pretko2018symmetry}. The relation can either be argued from a duality point of view \cite{pretko2018fracton}, where the gauge ``symmetry'' emerges from solving the momentum conservation equation, or starting with the observation that the gauge transformations in symmetric tensor gauge theories are identical to linearized diffeomorphisms, which is a symmetry in theories of elastic defects. Under this correspondence the immobile fractons map to disclinations, while the partially mobile fracton dipoles map onto dislocations (which satisfy the glide constraint). We note in passing that the duality in the context of elasticity has been previously studied in a series of papers by Kleinert\cite{kleinert1982duality,kleinert1983dual}, where (Euclidean) symmetric tensor gauge theories (of vector charge type) were introduced, however the glide constraint was omitted. It was further noted in [\onlinecite{gromov2017fractional}] (and later extended in Ref. [\onlinecite{slagle2018symmetric}]) that symmetric tensor gauge theories cannot remain gauge invariant in a general curved background, which is in striking difference compared to tranditional electrodynamics. This observation is in correspondance with a series of works [\onlinecite{slagle2017quantum, slagleXcube2017, slagle2018PRX,shirley2018foliated}], where the effective theory of the X-cube  model (a particular representative of type-I fracton models) was derived from the microscopics, studied in the presence of disclinations, and generalized to be defined on arbitrary foliated manifolds. The physical properties of these models (such as groundstate degeneracy and restricted mobility) depend on the geometry of the underlying space, thus we find it to be more appropriate to refer to fracton systems as geometric order, as suggested in [\onlinecite{slagleXcube2017}].

In yet another parallel development it was explained that the fracton phases can be further obtained by ``gauging'' subsystem symmetry \cite{williamson2016fractal,devakul2018fractal,you2018subsystem, williamson2018fractonic}. This leads to a swarm of microscopic models. At the same time, the long-distance description of the subsystem symmetries is not well-understood. A close relative of such symmetries, known as \emph{sliding} symmetry, appears in a theory of smectics (as well as other layered phases) that are studied in soft condensed matter physics \cite{lawler2004quantum, nussinov2005discrete, nussinov2005discrete}. On a lattice it is possible to define a subsystem symmetry that acts on a fractal set of lattice sitee. Gaugin such symmetry leads to the type-II fracton phases\cite{williamson2016fractal, devakul2018fractal}. Parity breaking phases of fractons, with possible gapless boundary modes were studied in [\onlinecite{you2018symmetric}, \onlinecite{prem2017emergent}, \onlinecite{gromov2017fractional}]. Further work on fracton phases and related topics can be found in [\onlinecite{prem2018cage, ma2017fracton, ma2017topological, petrova2017simple, yan2018fracton, pai2018localization, prem2018pinch, devakul2018correlation, devakul2018fractal2, pretko2017higher, bulmash2018braiding, slagle2018foliated}], while a broad picture is given in the review [\onlinecite{nandkishore2018fractons}]. Finally, we note that a similar formalism has been used to describe the geometric properties of fractional quantum Hall states. Namely, the foliated spacetime has been used in [\onlinecite{son2013newton, son-torsion, gromov-thermal, bradlyn2014low}]  to describe the transport of energy and momentum, while the symmetric tensor degrees of freedom have appeared in the context of nematic quantum Hall states and collective magnetoroton modes [\onlinecite{2011-Haldane-FQHE, you2014theory, maciejko2013field, golkar2013spectral, golkar2016higher, gromov2017Bimetric, nguyen2018fractional}].

\subsection{Summary of results}

The main objective of the present manuscript is to introduce a language that allows to \emph{systematically} construct a rich variety of effective field theories, which exhibit the phenomenology of both type-I and type-II fracton models. This will be done in such a way that both types appear as particular cases of the general framework.

Our construction rests upon an extension of a space(-time) symmetry algebra, which we dub the \emph{multipole algebra}. This algebra is a natural generalization of the symmetry algebras generated by the polynomial shift symmetries studied in\,[\onlinecite{nicolis2009galileon,griffin2013multicritical,hinterbichler2014goldstones, griffin2015scalar}]. These symmetries were originally introduced in the context of Galileon gravity\cite{nicolis2009galileon}. In systems with a conserved $U(1)$ charge these global symmetries lead to the conservation of the (various components of) multipole moments of the charge density. The aforementioned symmetries \emph{cannot} be regarded as ``internal'' because they do not commute with spatial translations and rotations. Throughout the manuscript the symmetry generators will be denoted as $\mathcal P^I_a$. These algebras are, in general, quite delicate objects as certain consistency conditions must be satisfied. These conditions arise from the intricate interplay between the spatial and multipole symmetries. If the multipole generators are picked ``at random'' the algebra will only close if all spatial symmetries are discarded, leading to trivial theories.

After defining the multipole algebra we explain how to construct effective field  theories, invariant under its action. We will restrict our attention to the matter described either by a real scalar ``phase'' field or by a charged complex scalar. These theories are introduced by first constructing \emph{all} possible invariant derivative operators, consistent with the multipole algebra and then including all terms, allowed by the symmetries, in the effective action. Such theories usually break spatial rotations, have quite unusual scaling properties, and are generally quite exotic. Further, we discover that some of these theories in $d=3$ exhibit an enhanced sliding symmetry, alluded to in the Introduction. 

Gauging of the multipole algebra should lead to exotic theories of elasticity and/or gravity, because in order to consistently gauge these theories one \emph{must} gauge the spatial rotations and translations. This explains why symmetric tensor gauge theories (which are a particular case of the present construction) are very sensitive to the background geometry. It is, however, possible to ``partially'' gauge these symmetry algebras, under the assumption that the space(-time) part of the curvature tensor is trivial (simply put, in flat space, without torsion). This partial gauging procedure leads to a very rich set of gauge theories, which includes all known symmetric tensor gauge theories (and various variations thereof: higher rank; scalar, vector or tensor charge; tracefull/traceless, etc.) as well as the ``generalized gauge theories'' introduced in [\onlinecite{bulmash2018generalized}]. These gauge theories naturally satisfy the exotic Gauss law constraints, which in the present formalism, is \emph{systematically derived} by gauging the multipole algebra. These Gauss law constraints can be visualized, upon discretizing on a lattice, as prescribing the ``allowed'' charge configurations. These charge configurations specify which excitations are mobile, which are sub-dimensional and which are fractal. To be concrete, we derive the continuous model for the $U(1)$ version of the Haah code [\onlinecite{bulmash2018generalized}] from the symmetry principles.

The gauge theories described above are gapless and do not correspond to the gapped fracton phases. To further advance our construction we show, building upon [\onlinecite{BB2018Higgs}], that an extremely rich variety of phases emerges after the condensation of charge $k$ objects, which leads to the reduction in symmetry from $U(1)$ to $\mathbb Z_k$. It is particularly interesting that depending on value of $k$ some of the ``allowed'' charge configurations become redundant, while in other cases a complex charge configuration turns into a hopping operator, which hops over several lattice spacings. Using this procedure we find a version of the Sierpinski triangle models in two dimensions and the $\mathbb Z_2$ Haah code. We then explain how to translate the obtained results into the language of polynomials over finite fields. We further explain how to define multipole moments over a finite field directly from the polynomials. 

The paper is organized as follows. In Section II we introduce the generalized polynomial shift symmetries and use these symmetries to motivate the multipole algebra. Next we define the multipole algebra in abstract terms and illustrate the definition on a couple of examples. In Section III we explain how to construct the invariant field theories and explain how to partially gauge the multipole algebra, directly at the level of the matter theory. We investigate several examples of such gauge theories in two and three dimensions. In Section IV we discuss various extensions of the multipole algebra, most notably the charge condensation and crystalline symmetries. We also explain the relation between the present formalism and the approach based on the polynomials over finite fields. Finally, in Section V we present our conclusions and discuss open directions.

\section{Multipole Algebra} 

\subsection{Polynomial shift symmetry} 
The conservation of the dipole moment and its relevance to the fracton order was emphasized in\,[\onlinecite{pretko2017subdimensional},\onlinecite{pretko2017higher}]. The symmetries that have to do with the conservation of the multipole moments have been extensively studied prior to these works\,\cite{nicolis2009galileon,griffin2013multicritical,hinterbichler2014goldstones, griffin2015scalar} and are known as \emph{polynomial} shift symmetries. To simplify the presentation we will first introduce these symmetries by specifying the action on the matter fields. To be concrete, consider a real scalar field $\varphi$. The action of the polynomial shift symmetry is defined according to
\be
\la{eq:Polyshift}
\delta \varphi = \lambda_{\alpha} P^\alpha(x) \,,
\ee
where $\lambda^\alpha$ is a symmetry parameter and $P^\alpha(x)$ is an arbitrary polynomial; the sum over $\alpha$ is understood. We will further assume that there is a \emph{finite} number of transformations that appear in \eqref{eq:Polyshift}. If the dynamics of $\varphi$ is described by an action $S[\varphi]$, which is invariant under \eqref{eq:Polyshift}, then we have the following set of conserved charges
\be\la{eq:Charges}
 Q^\alpha = \int d^dx q^\alpha(x)\,,
\ee
where $q^\alpha(x)$ is the charge density that can be found via a direct application of Noether's theorem. If among $P^\alpha(x)$ there is \emph{constant} polynomial -- corresponding to the global $U(1)$ charge conservation -- then we can write a more intuitive expression for the charges. Denoting the charge density as $\rho(x)$ we find
\be\la{eq:Charges2}
 Q^\alpha = \int d^dx P^\alpha(x) \rho(x)\,.
\ee
Which implies the conservation of various generalized multipole moments. If we further assume that the polynomials $P^\alpha$ are homogeneous
\be\la{eq:homogeneous}
P^{I_a}_{a}(x)= \sum_{i_1,i_2,\ldots,i_a} \mu^{I_a}_{i_1i_2\ldots i_a}x^{i_1}x^{i_2} \ldots x^{i_a}\,,
\ee
we can identify \eqref{eq:Charges2} with the (components of) proper multipole moments. To be specific,
\be\la{eq:Charges3}
Q^{I_a} = \int d^dx \,\, \mu^{I_a}_{i_1i_2\ldots i_a}x^{i_1}x^{i_2} \ldots x^{i_a} \rho(x)\,.
\ee

Notice, that since the coordinates $x$ are dimensionalfull, then so are the parameters $\lambda_{\alpha}$. This ultimately leads to the generalized Mermin-Wagner theorem which states that if the power of the polynomials in \eqref{eq:Polyshift} is no larger than $n$, then the symmetry cannot be spontaneously broken in $d\leq n + 1$ spatial dimensions \cite{griffin2015scalar}. This is quite similar to the Mermin-Wagner theorem for the higher form symmetry \cite{lake2018higher}. The relation between these two types of symmetry warrants further exploration.

A particularly simple case of this structure is the symmetry under \emph{all} polynomial shifts of the degree no greater than $n$. Such transformation takes form 
\be \la{eq:Symmetric}
\delta \varphi = \lambda+\lambda^{(1)}_i x^i + \lambda^{(2)}_{ij}x^{j}x^j +\ldots\,.
\ee
This leads to the conservation of \emph{all} multipole moments of degree less or equal to $n$. The conserved charges are the arbitrary moments of the density $Q^{ij\ldots}=\int d^dx \,\rho(x) x^i x^j\ldots$ The case of $n=1$ corresponds to the conservation of the dipole moment and leads to the scalar charge theory \cite{pretko2017subdimensional}, while the restricted mobility of dipoles can be added by supplementing the $n=1$ symmetry with $\delta \varphi = \lambda^\prime|x|^2$, which leads to the \emph{traceless} scalar charge theory. 

\subsection{Multipole algebra} 

The transformations \eqref{eq:Polyshift} commute with each other and form an unremarkable algebraic structure. However, these transformations \emph{do not commute} with the spatial symmetries: translations and rotations. Instead, the polynomial shift symmetries extend the algebra of spatial symmetries to a bigger \emph{multipole algebra}, $\mathfrak{m}$. Consequently, the symmetries responsible for the conservation of the multipole moments are \emph{not internal}. This provides a general explanation to the observation made in Ref.\,[\onlinecite{gromov2017fractional}] (and later extended in [\onlinecite{slagle2018symmetric}]) that the symmetric tensor gauge symmetry is ``broken'' on a general curved manifold.

\subsubsection{Intuitive preamble}

Before diving into the formal details of the multipole algebra we would like to illustrate the physical origins of the structure. Consider a transformation law of a quadrupole moment of the charge density under a translation by the vector $r^k$
\be\la{eq:translationD}
\delta Q^{ij} = r^j Q^i + r^i Q^j + r^i r^j Q\,,
\ee
where $Q^i$ is the total dipole moment and $Q$ is the total charge. Usually, one would state that provided that all lower moments vanish, the quadrupole moment is invariant under translations. Imagine a situation when $Q^1=Q^2=0$ enforced by the symmetry and all other components of the dipole moment are non-zero. Then the quadrupole moment is invariant under translations in the $x_1-x_2$ plane and we can further constrain it by the symmetries in within this plane. Effective description of systems with such constraints will be translation invariant in $x_1-x_2$ plane only.

We will further encounter another interesting degenerate case, which will play the central role in the discussion of the fractal phases. Consider a particular component of the quadrupole moment $Q_\mu = Q^{ij}\mu_{ij}$, where $\mu_{ij}$ is a certain degenerate matrix with the property that its kernel equals to the orthogonal complement of the $x_1-x_2$ plane. Then such component of the quadrupole moment is translationally invariant under \emph{all} translations
\be
\delta Q_\mu = \mu_{ij}r^j Q^i  + \mu_{ij}r^i Q^j + \mu_{ij}r^ir^j Q=0\,. 
\ee
This equality holds for \emph{all} translations since if we take translations outside of the $x_1-x_2$ plane the matrices $\mu_{ij}$ will project them down to the $x_1-x_2$ plane, but in this plane the projection of the total dipole moment vanishes. Thus $Q_\mu$ is translation invariant. 

The multipole algebra formalizes these simple ideas in the language of symmetry.

\subsubsection{Formal definition of the multipole algebra}

To get some insight into the algebraic structure we need to sort the polynomials by their degree. We introduce a set of generators of the polynomial symmetries, $\mathcal P^{I_a}_a$, so that $a$ corresponds to the degree of the polynomial, whereas $I_a$ runs through all polynomials of degree $a$. Then the general multipole algebra is defined via the following set of commutation relations
\bea\la{eq:Multi}
&&\!\!\!\!\!\![R_{ij}, T_k]=\delta_{k[i}T_{j]}\,, \quad [R_{ij}, R_{kl}]=\delta_{[k[i}R_{j]l]}\,,
\\\la{eq:MultipoleAlg}
&&\!\!\!\!\!\![R_{ij}, \mathcal P_a^{I_a}] = f_{ij}{}^{I_a}{}_a{}^b{}_{J_b} \mathcal P_b^{J_b},\quad [T_i, \mathcal P_a^{I_a}] = f_i{}^{I_a}{}_a{}^b{}_{J_b} \mathcal P_{b}^{J_b}\,,
\eea
where $O_{[ij]}$ denotes anti-symmetrization over $i,j$ and $f_{ij}{}^{I_a}{}_a{}^b{}_{J_b}, f_i{}^{I_a}{}_a{}^b{}_{J_b}$ are the structure constants, that define the algebra. In the polynomial representation the content of \eqref{eq:MultipoleAlg} is quite simple: rotations and translations, when applied to the polynomials, should produce linear combinations of the polynomials within the multipole algebra, \emph{i.e.} the set of polynomials $\mathcal P_a^{I_a}$ is \emph{closed} under rotations and translations. The multipole algebra can be exponentiated to  a multipole group, $\mathfrak M$. The multipole group is, in a way, a more fundamental construct since, unlike the algebra, it admits a crystalline analogue and survives the condensation of charge $p$ objects (provided the charge is well-defined) as a discrete group.

 If the polynomials, in the polynomial representation, depend on $d$ variables, then the indices $i,j$ run from $1$ to $k\leq d$. This happens because generally the vector space of polynomials corresponding to $\mathcal P_a^{I_a}$ is not compatible with translations or rotations; in such case the multipole algebra is trivial. However, if $\mathcal P_a^{I_a}$ are chosen carefully then very rich and intricate structures can emerge.

\subsection{Maximal multipole algebra} 

 Next we turn to consider a few explicit examples. The simplest case is the set of transformations \eqref{eq:Symmetric} with $n=1$. It gives rise to the following algebra (we omit the usual relations between translations and rotations \eqref{eq:Multi})
\be\la{eq:MAlgebra1}
[T_i, \mathcal P_1^j]= \delta_{ij}\mathcal P^0\,, \qquad [R_{ij}, \mathcal P_1^k]=\delta_{k[i}\mathcal P_1^{j]}\,,
\ee
where $\mathcal P^0$ refers to the constant shifts. We emphasize that the internal index $I$ was identified with the spatial index $i$. This is due to an ``accidental'' isomorphism between the set of linear shifts and set of translations. One distinguishing feature of this algebra is that is it consistent with all spatial translations and rotations. This happens because all polynomials of degree $\leq 1$ were included, consequently such set is closed under any operation that does not increase the power of the polynomial.

One simple extension of the algebra \eqref{eq:MAlgebra1}, consistent with all spatial symmetries, is the addition of one extra generator $\mathcal P_2^0$, corresponding to $\delta \varphi = \lambda^\prime |x|^2$. This generator leads to the commutation relations
\be\la{eq:MAlgebra2}
[T_i,\mathcal{P}_2^0]=\mathcal P_1^i\,, \qquad [R_{ij}, \mathcal{P}_2^0]=0\,.
\ee
We note in passing that the generators $\{T_i,R_{ij},\mathcal{P}^0, \mathcal{P}_1^i, \mathcal{P}_2^0\}$ form the Bargmann algebra\cite{andringa2011newtonian}, where the spatial translations $T_i$ correspond to the Galilean boosts, linear shifts $\mathcal{P}_1^i$ correspond to the spatial translations, $\mathcal{P}_2^0$ to the Hamiltonian and $\mathcal{P}^0$ to the mass central charge. It would be interesting to explore this isomorphism to construct field theories invariant under this type of multipole algebra.

In the above set of examples it is amusing to note that the commutation relations between the $R_{ij}$ and $\mathcal P_1^k$ are the same as between $R_{ij}$ and $T_k$. In the absence of the generator $\mathcal P_2^0$, which provides the asymmetry, we could swap $T_k$ and $\mathcal P_1^k$ and obtain the same algebra back. This observation has a nice elastic interpretation. If the theory of elasticity is viewed as a gauge theory of translations and rotations, $T_d\rtimes SO(d)$, then we could use either real translations \emph{or} shift symmetries to construct such a gauge theory. This ambiguity corresponds to the two distinct approaches to the relationship between fractons and elasticity \cite{pretko2018fracton,gromov2017fractional}. If the generator $\mathcal P^2_0$ is introduced \emph{and} gauged, then one will end up with two different theories. We will elaborate on this distinction in a forthcoming work.

These symmetric cases can be generalized to arbitrary multipole moments. In order to include the multipole moments up to $n$ in $d$ spatial dimensions we introduce a set of polynomial symmetries described by arbitrary symmetric tensors of degree $a$, $\mathcal P_a^{I_a} = \mathcal P_a^{i_1\,\ldots i_{a}}$. Each such tensor has $\binom{d+n-1}{n}$ independent components. Together with translations and rotations these symmetry generators form the following algebra
\bea\la{eq:multipolen}
[T_j, \mathcal P_a^{i_1i_2\ldots i_a}] &=& \mathcal P_{a-1}^{i_1\ldots i_{m-1} i_{m+1}\ldots i_a}\,, \quad \text{if}\,\, j=i_m
\\ \nonumber
&=& 0 \qquad \qquad\qquad  \qquad\,\, \text{otherwise}
\\
\la{eq:multipolenR}
[R_{jl},\mathcal P_a^{i_1i_2\ldots i_a}] &=& \delta_{[j[i_1}\mathcal P_a^{l]i_2]\ldots i_a}\,, \,\,\, \text{if} \,\, j=i_m\,, l=i_{m^\prime}
\\\nonumber
&=& 0 \qquad \qquad\qquad\,  \text{otherwise}\,.
\eea
Theories that are symmetric under this algebra conserve all the multipole moments up to order $n$. Such multipole algebras are completely characterized by the spatial dimension and a single integer $n$, thus we will refer to those as \emph{maximal} multipole algebra of order $n$, $\mathfrak m_{\rm max}^n$. It may be tempting to speculate \cite{pretko2017subdimensional} that the theory obtained by gauging the algebra \eqref{eq:multipolen} (perhaps, in the limit $n\rightarrow \infty$), may be related to the Vassiliev theories of higher spin gravity. In view of the algebra \eqref{eq:multipolen}-\eqref{eq:multipolenR}, this speculation seems unlikely since the symmetric multipole generators all commute with each other, whereas in the Vassiliev higher spin theory they form a version of the $W_\infty$ algebra (also referred to as $shs(1)$ in earlier works \cite{fradkin1987candidate}).

\subsection{Homogeneous multipole algebra} 

Next we consider a less symmetric, but still very palatable case. Consider a set of polynomial symmetries where all polynomials $P^{I_a}_a$ are homogeneous, that is of the form \eqref{eq:homogeneous}. We will refer to such multipole algebras as \emph{homogeneous}. We would like to demonstrate how certain rotation and translation generators drop form the algebra due to requirement that the set of polynomials $P^{I_a}_a$ is closed under as many translations and rotations as possible. Consider, for example, 
\be\la{eq:Example}
\delta \varphi = \sum_{a,I_a} \lambda_{I_a} P^{I_a}_a(x)\,, 
\ee
where the five polynomials $P^{I_a}_{a}(x)$ are given by
\bea\la{eq:mu1}
&&P^0_0 = 1\,, \,\,\, P^1_1=\mu^1_i x^i\,, \,\,\, P_1^2 =\mu^2_i x^i
\\\la{eq:mu2}
&& P_2^1 = \mu^1_{ij}x^ix^j\,, \,\,\, P^2_2 = \mu^2_{ij}x^ix^j \,,
\eea
where $\mu^{I_1}_{i}$ and $\mu^{I_2}_{ij}$ are \emph{fixed} rank-1 and rank-2 tensors and $i,j=1,\ldots,d$. These tensors are fixed up to the freedom of replacing $\mu^{I_1}_i$ and $\mu^{I_2}_{ij}$ with the linear combinations thereof. 

 Depending on the circumstances, these polynomials can give rise to several multipole algebras. First, assume that $\mu^{I_2}_{ij}$ are all non-degenerate. Then only two out of $d$ translations survive. These are translations in the directions $t^j_{(1)}$ and $t^j_{(2)}$, which are explicitly determined from 
\bea\la{eq:trans1}
&&\mu^1_{ij}t_{(1)}^j = \alpha_1\mu^1_i + \alpha_2\mu^2_i\,,
\\\la{eq:trans2}
&&\mu^2_{ij}t_{(2)}^j = \beta_1\mu^1_i + \beta_2\mu^2_i\,,
\eea
where $\alpha_i,\beta_i$ are some constants. Eqs.\eqref{eq:trans1}-\eqref{eq:trans2}

Next we consider rotations. Clearly, the only rotation generator that has a chance to survive is the rotation in $\mu^1_i-\mu^2_j$ plane, $R_{12}$. Within this plane we can always arrange that 
\be
[R_{12},\mu^1_{ij}] = \mu^2_{ij}\,, \qquad [R_{12},\mu^1_{ij}] = -\mu^2_{ij}\,,
\ee
via redefining the symmetry transformations in such a way that, within the $\mu^1_i-\mu^2_j$ plane, $\mu^1_{ij}\propto \sigma^1$ and $\mu^2_{ij}\propto \sigma^3$, where $\sigma^i$ are the Pauli matrices. Thus we obtain the multipole algebra
\bea\la{eq:multiexample1}
&&[T_1, \mathcal P_1^{I_1}] = [T_2, \mathcal P_1^{I_1}] = \mathcal P^0_0\,,
\\\la{eq:multiexample2}
&& [T_1, \mathcal P_2^{I_2}] =  f_1{}_{I_1}^{I_2}\mathcal P_1^{I_1}\,, \qquad [T_2, \mathcal P_2^{I_2}] =  f_2{}_{I_1}^{I_2}\mathcal P_1^{I_1}
\\\la{eq:multiexample3}
&& [R_{12}, \mathcal P_2^{I_2}]=\epsilon^{I_2}{}_{J_2} \mathcal P_2^{J_2}\,,
\eea
where sum over repeated $I_a,J_a$ is understood. Although, we have started in $d$ spatial dimensions, only two translation generators and one rotation generator have survived.

There is another interesting possibility, which will arise in the study of the $U(1)$ Haah code. When both $\mu^{I_2}_{ij}$ are degenerate in such a way that the kernels of $\mu^{I_2}_{ij}$ coincide with each other and with the orthogonal complement of $\mu^{I_1}_i$. Then Eqs.\eqref{eq:trans1}-\eqref{eq:trans2} hold true for \emph{all} translations since $\mu^{I_2}_{ij}$ are projectors to the $\mu^1_i-\mu^2_i$ plane. In such algebra we get an additional set of trivial commutation relations
\be\la{eq:multiexample4}
[T_i, \mathcal P_2^{I_2}] = 0\,,
\ee
where $i$ runs over the orthogonal complement to $\mu^{I_1}_i$: from $1$ to $d-2$ in the present example.

\section{Invariant field theories} 

\subsection{General constraints} 

We turn to the construction of the field theories invariant under the action of the multipole algebra. First, we fix a multipole algebra $\mathfrak m$ and construct an irreducible representation. To start, we have to fix the transformation law under rotations. For simplicity we take a single real scalar field $\varphi$\,\footnote{This choice excludes the vector charge theories. The latter are obtained by choosing $\varphi$ to be a vector  under the rotations}. The transformation laws under the action of $\mathcal P_a^{I_a}$ are given by Eq. \eqref{eq:Polyshift}. We will denote the highest power that appears in \eqref{eq:Polyshift} as $a_{\rm max}$. The time derivative term will take the ordinary form, $\dot{\varphi}\dot{\varphi}$\,\footnote{This is so because we are considering the shift symmetries, which are polynomial in the spatial coordinates, and not in time.}. To construct the kinetic term, we need an \emph{invariant} derivative operator, \emph{i.e.} a derivative operator, consistent with \eqref{eq:Polyshift}. To this end we consider a general differential operator
\be\la{eq:Ddef}
D = q + q^{i_1}\p_{i_1} + q^{i_1i_2}\p_{i_1}\p_{i_2}+\ldots+q^{i_1\ldots i_s}\p_{i_1}\ldots \p_{i_s},
\ee
where $s\leq a_{\rm max}$. The coefficients $q^{i_1i_2\ldots}$ will be chosen in such a way that $D \varphi$ is invariant under the action of $\mathcal P_a^{I_a}$. In the most general case these equations take form
\be\la{eq:general}
D P_a^{I_a} ={0}\,,
\ee
where $P_a^{I_a}$ is the polynomial corresponding to the action of $\mathcal P_a^{I_a}$. Eq.\eqref{eq:general} must hold for \emph{all} $P_a^{I_a}$. The solutions to these equations are the differential operators $D_\alpha$ (the index $\alpha$ will label the solutions). The constraints \eqref{eq:general} can be written more explicitly if we introduce the following parametrization for $P_a^{I_a}(x)$
\be
P_a^{I_a}(x)= \mu^{I_a} + \mu^{I_a}_{i_1}x^{i_1}+\ldots+\mu^{I_a}_{i_1\ldots i_a}x^{i_1}\ldots x^{i_a}.
\ee
Then Eq.\eqref{eq:general} turns into a set of \emph{linear} equations on $q^{i_1\ldots}$
\bea\la{eq:constr1}
&& q\mu^{I_0} + q^{i}\mu^{I_1}_{i} + q^{ij} \mu^{I_2}_{ij}+\ldots=0\,,
\\\la{eq:constr2}
&& q\mu^{I_1}_{i_1} + q^{i}\mu^{I_2}_{i_1i} + q^{ij}\mu^{I_3}_{i_1ij}+\ldots=0\,,
\\\nonumber
&& \ldots\,,
\\\la{eq:constrn1}
&& q \mu^{I_{a-1}}_{i_1\ldots i_{a-1}} + q^{i}\mu^{I_a}_{i_1\ldots i_a} = 0\,,
\\\la{eq:constrn}
&&q \mu^{I_a}_{i_1\ldots i_a} = 0\,,
\eea
which hold for \emph{all} $I_a$.
It immediately follows that $q=0$. Thus, the invariant derivatives must start with at least $q^{i_1}\p_{i_1}$. This implies an additional invariance of the effective theory under a global $U(1)$ transformation $\delta \varphi = \mu^{I_0}$. While the multipole algebra does not require to have a constant shift as a separate generator, it seems difficult to find a matter theory that represents such algebra polynomially without a symmetry enhancement. 

The system \eqref{eq:constr1}-\eqref{eq:constrn} contains $\frac{(2a)!}{a!}$ equations for \emph{every} polynomial, and only $\frac{(2a)!}{a!}$ unknowns. Thus, generally it is severely overdetermined and has no solutions. As we will see shortly, there are, indeed, ``degenerate'' cases when the system does admit solutions. This phenomenon is somewhat reminiscent to the existence of a solution to the (severely overdetermined) pentagon and hexagon equations, which are the consistency conditions for fusion tensor categories.

If there are no solutions for $s\leq a_{\rm max}$ then the system can \emph{always} be solved by higher order differential operator that annihilates \emph{all} polynomials of degree no grater than $a_{max}$. Such operator takes form
\be\la{eq:D2ap1}
D= q^{i_1\ldots i_{a_{\rm max}+1}}\p_{i_1}\ldots \p_{i_{a_{\rm max}+1}}\,,
\ee
for \emph{any} $q^{i_1\ldots i_{a_{\rm max}+1}}$. The action constructed using these solutions will have an enhanced symmetry under \emph{all} polynomial shifts and will represent the maximal multipole algebra $\mathfrak m_{\rm max}^{a_{\rm max}}$. Presently it is not clear how to establish the existence of a solution to \eqref{eq:constr1}-\eqref{eq:constrn} without solving $O(4^a a^a)$ equations. 

A comment is in order. In the above construction, as well as in the remainder of the manuscript, we demand the complete invariance of the action under the symmetries. It is, however, possible to consider a weaker condition -- the invariance of the action up to a total derivative. This condition allows for a wider array of the invariant Lagrangians. We leave exploration of this scenario for the future work.

\subsection{Constraints in the homogeneous case} 

In the remainder of the manuscript we will focus on the homogeneous multipole algebras. The constraint equations take form
\bea \la{eq:homoConst}
&&\mu^{I}_i q^i=0\,,\ 
\\ \la{eq:homoConst2}
&&\mu^{I}_{ij}q^{i}=0\,, \qquad \mu^{I}_{ij}q^{ij}=0\,,
\\\la{eq:tric}
&&\mu^{I}_{ijk}q^i=0\,,\quad \mu^{I}_{ijk}q^{ij}=0\,,\quad \mu^{I}_{ijk}q^{ijk}=0\,,
\\ \nonumber
&&\ldots.
\eea
The system of equations \eqref{eq:homoConst}-\eqref{eq:tric} is more overdetermined than \eqref{eq:constr1}-\eqref{eq:constrn}, since the contractions such as $\mu^{I}_{ijk}q^{ij}=0$ must vanish separately. Nevertheless, as we will see shortly, these systems still admit solutions.

Finally, we note that the solutions of \eqref{eq:homoConst}-\eqref{eq:tric} are determined \emph{up to an overall scale} (which is not the case for \eqref{eq:constr1}-\eqref{eq:constrn}). We will discuss how to fix the scale later.

To the lowest order in derivatives, the invariant effective action is 
\be\la{eq:InvAction}
S = \int d^dxdt \Big[\dot \varphi\dot \varphi - \sum_{\alpha,\beta}\lambda_{\alpha\beta} (D_\alpha \varphi)(D_\beta \varphi) \Big]\,,
\ee
where $\alpha$ runs over \emph{all} solutions of \eqref{eq:general}. 

To the lowest order in gradients the effective action is quadratic in the invariant derivatives. However, the latter are not necessarily of the same degree. The derivative expansion of such theories is quite unusual. In more traditional situations global symmetries restrict the terms that appear in the effective action to \emph{all} orders in derivative expansion. In the present case the situation is drastically different. To the lowest order(s) in we have a few special derivative operators $D_\alpha$. The leading terms in the gradient expansion must be written using $D_\alpha$ operators only. Such terms dominate up to the order $2a_{\rm max}$. However, when we move further in the gradient expansion, say to the order $2a_{\rm max}+1$, we are allowed to use \emph{any} operator of the form \eqref{eq:D2ap1}. 


In the case of the maximal multipole algebra $\mathfrak m^n_{\rm max}$ it is possible to write a rotationally-invariant action
\be\la{eq:Horava}
S = \int d^dxdt \Big[\dot \varphi\dot \varphi - \sum_{k> 1} \lambda_k (D_{i_1i_2\ldots i_n} \varphi D^{i_1i_2\ldots i_n} \varphi)^k \Big]\,,
\ee
where $D_{i_1i_2\ldots i_n} = \p_{i_1}\ldots\p_{i_n}$. Rotationally invariant actions of the type \eqref{eq:Horava} were studied in [\onlinecite{nicolis2009galileon,griffin2013multicritical,hinterbichler2014goldstones, griffin2015scalar}]. We will view the actions \eqref{eq:InvAction}-\eqref{eq:Horava} within the framework of the effective field theory. In particular, this entails to including \emph{all} possible terms, consistent with the postulated symmetries and organizing these terms according to some degree of ``relevance''.

\subsection{Multipole gauge theory} 

In this Section we will explain how to gauge the multipole symmetries.
With the invariant derivatives at hand it is now possible to introduce a local version of the multipole symmetry. In principle, we could do it directly from the commutation relations, however this entails to gauging the spatial symmetries as well. We will leave this program to a future work.

Present approach allows to ``partially gauge'' the symmetries: the multipole symmetries will become local, while rotations and translations will remain global. This amounts to setting the corresponding gauge fields to $0$, so that Ricci curvature and torsion vanish. The gauging procedure is well known -- we promote the global symmetry in \eqref{eq:Polyshift} to a local one $\delta \varphi = \zeta(x)$ and introduce a \emph{covariant} derivative operator
\be
\nabla_{\alpha} \varphi = D_{\alpha}\varphi + a_{\alpha}(t,x)\,,\quad  \delta a_{\alpha}(t,x) = - D_{\alpha} \zeta\,.
\ee
where $a_{\alpha}(t,x)$ is the gauge field. It is importnant to note here that $a_{\alpha}$ is the space-time scalar, in that it does not transform under rotations or translations, provided the coefficients $q^{i_1\ldots}$ transform as proper tensors. The time derivative is replaced with the ordinary $\nabla_0 \varphi = \dot{\varphi} + \chi$\,, with $\delta \chi = -\p_0\zeta$.

With the gauge fields at hand we introduce a set of conjugate momenta (or, ``electric fields'') according to
\be
\Big[e_{\beta}(x^\prime),a_{\alpha}(x)\Big]=-i \delta(x-x^\prime) \delta_{\alpha \beta}\,.
\ee
The curvature of the connection (or, ``magnetic field'') is guaranteed to be non-vanishing if we gauge the entire algebra $\mathfrak m$, provided that multipole constraints do not reduce the spatial symmetries to one or zero dimensions. In the latter case the curvature will vanish identically. In present approach we will construct the magnetic fields on the case-by-case basis as as gauge-invariant combinations of $a_\alpha$. 

The general invariant Lagrangian takes form
\be\la{eq:InvLag}
\mathcal L = \mathcal L[\nabla_{\alpha} \varphi] + \sum_{\alpha} e_{\alpha}\dot{a}_{\alpha} - \mathcal H[e,b]\,,
\ee
where $\mathcal H[e,b]$ schematically denotes the Hamiltonian for the gauge fields. The action is supplemented with the Gauss law constraint (obtained by integrating out $\chi$)
\be\la{eq:Gauss}
D^\dag_{\alpha} e_{\alpha} = \rho\,,
\ee
where the conjugate derivative $D^\dag_{\alpha}$ is defined as
\be
\int d^dx dt f (D_{\alpha} g) = \int d^dx dt (D^\dag_{\alpha}f)g\,.
\ee
The Gauss law of the type \eqref{eq:Gauss} was postulated in Ref.\,[\onlinecite{bulmash2018generalized}]. We find that \eqref{eq:Gauss} follows directly from the underlying structure of the multipole algebra. At the same time, not \emph{every} Gauss law constraint is consistent with some multipole algebra.

The invariant derivatives $D_\alpha$ can be discretized on a lattice, leading to various charge configurations. To be concrete (see Ref.\,[\onlinecite{bulmash2018generalized}]), consider an operator $e^{-ia_\alpha(x)}$, acting on a site with label $x$. This operator changes the value of the electric field $e_\alpha(x)$ at the same lattice site, say raises it by $1$. This, in turn, requires to introduce electric charges at all lattice sites $x^\prime$, that are connected to $e_\alpha(x)$ by the Gauss law \eqref{eq:Gauss}. We will make heavy use of this pictorial representation.

Such configurations of finite number of point charges are characterized by a set of multipole moments. These moments are determined by $q^i_{\alpha}, q^{ij}_{\alpha},\ldots$. Only the lowest of these moments is independent of the choice of coordinate origin, however the solution of \eqref{eq:homoConst}-\eqref{eq:tric} is independent of the choice of origin. Indeed, assume that the total charge is zero, then the dipole moment is well-defined. When the origin is shifted by $r^i$ the quadrupole moment transforms as $\delta q^{ij}_{\alpha} = r^i q^j_{\alpha} + r^j q^i_{\alpha}$, which still satisfies \eqref{eq:homoConst}-\eqref{eq:tric}. In the most general, non-degenerate case the solutions are invariant only under the translations in ``allowed'' directions, such as the ones specified by \eqref{eq:trans1}-\eqref{eq:trans2}. This is not too surprising since other translations do not belong to the symmetry algebra.

\subsection{Maximally symmetric gauge theory} 

Next we consider a few examples of the general formalism outlined above. First, we would like to make sure that the symmetric tensor gauge theories follow. This is indeed so, provided we gauge the maximal algebra $\mathfrak m^n_{\rm max}$. The covariant derivative takes form
\be
\nabla_{i_1i_2\ldots i_{n+1}} \varphi = \p_{i_1}\p_{i_2}\ldots\p_{i_{n+1}} \varphi + a_{i_1i_2\ldots i_{n+1}}\,.
\ee
In this case a rotationally invariant action is possible
\be
S = \int d^dx dt \Big[\nabla_0\varphi\nabla_0\varphi - \nabla_{i_1i_2\ldots i_{n+1}} \varphi\nabla^{i_1i_2\ldots i_{n+1}} \varphi\Big]\,.
\ee
This type of (ungauged) actions has been studied in great detail in Refs.[\onlinecite{griffin2013multicritical,hinterbichler2014goldstones, griffin2015scalar}], in relation to ``slow'' Goldstone bosons. 

Including, additionally, pure trace generators of one higher degree $\mathcal P^{I_{n+1}}_{n+1}$ necessitates the change in the covariant derivative
\be
\nabla_{\alpha}\varphi = q_{\alpha}^{i_1i_2\ldots i_{n+1}}\p_{i_1}\p_{i_2}\ldots\p_{i_{n+1}} \varphi + a_{\alpha}\,,
\ee
where $q_{\alpha}^{i_1i_2\ldots i_n}$ are symmetric \emph{traceless} tensors and the gauge field $a_{\alpha}$ can be related to the symmetric tensor gauge field via projecting out the tracefull parts $a_{\alpha}=q_{\alpha}^{i_1i_2\ldots i_{n+1}} a_{i_1 i_2\ldots i_{n+1}}$. Both traceless and tracefull scalar charge theories exhibit a scaling symmetry with dynamical critical exponent $z=n+1$
\be
t\rightarrow \lambda^z t\,, \,\,\, x_i \rightarrow \lambda x_i\,,\,\,\, \varphi \rightarrow \lambda^{\frac{n+1-d}{2}}\varphi\,.
\ee
Thus, in $d$ spatial dimensions, the field $\varphi$ is dimensionless if $n+1=d$. As discussed in [\onlinecite{griffin2015scalar}], this corresponds to a generalized version of the Mermin-Wagner theorem: the (maximal) multipole symmetry of degree $n$ cannot be spontaneously broken in $d\leq n+1$ dimensions. 

\begin{figure}
\includegraphics[width=3.5in]{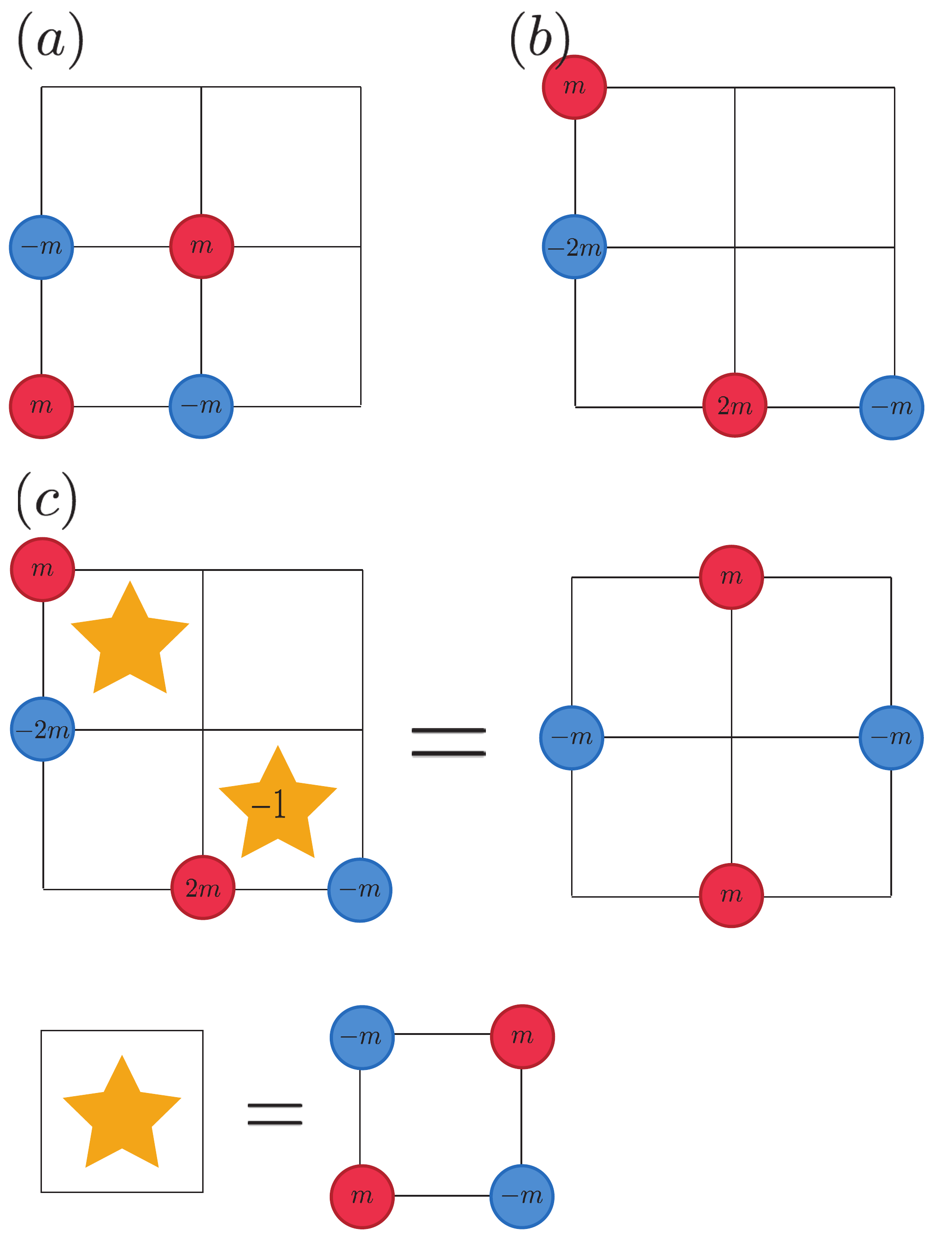}
\caption{(a) The charge configuration corresponding to $q^{ij}\propto\sigma^1$. (b) The charge configuration corresponding to $q^{ij}\propto\sigma^3$. (c) A more convenient basis of charge configurations is obtained by applying (a) and inverse of (a) at plaquettes labeled by star and $-1$ star correspondingly.}
\label{traceless}
\end{figure}

In the case $n=1$ we get the effective theory for the traceless scalar charge theories
\be
S = \int d^dx dt \Big[\nabla_0\varphi\nabla_0\varphi - (q^{i_1i_2}\nabla_{i_1i_2} \varphi) (q^{j_1j_2}\nabla_{j_1j_2} \varphi) + \ldots\Big]\,,
\ee
where $q^{ij}$ are symmetric traceless tensors. The charge configurations for these theories have been discussed previously\cite{pretko2017subdimensional}. We will use this case as the first benchmark of the present formalism. In two dimensions the Gauss law takes form 
\be
D^\dag_{1} e_{1} + D^\dag_{2} e_{2} = \rho\,,
\ee
since there are only two symmetric traceless tensors in $d=2$, namely the Pauli matrices $\sigma^1_{ij}$ and $\sigma^3_{ij}$. The corresponding charge configurations are illustrated in Fig.\ref{traceless}.

\subsection{Quadratic multipole algebras in two dimensions} 

\begin{figure*}
\includegraphics[width=\textwidth]{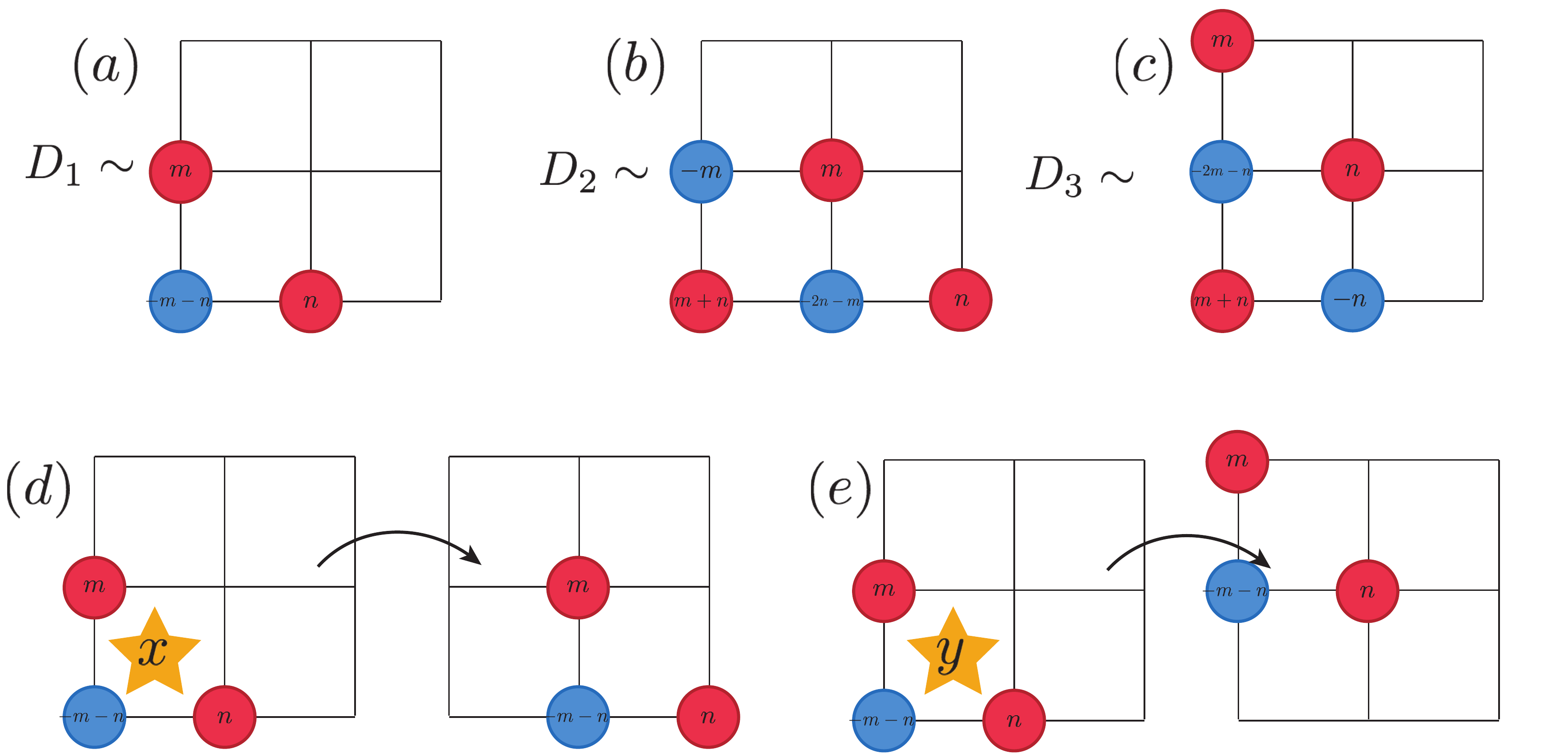}
\caption{(a)-(c) The elementary charge configurations, corresponding to $D_\alpha$, for the degenerate theory, characterized by \eqref{eq:MAdipole2D1}-\eqref{eq:MAdipole2D2}, with invariant derivatives given by \eqref{eq:invDdipole}.(d) Application of the charge configuration corresponding to $D_2$ results in hopping of the $(n,m)$ dipole in $x$-direction. (e) Application of the charge configuration corresponding to $D_3$ results in hopping of the $(n,m)$ dipole in $y$-direction.}
\label{Dipole2d}
\end{figure*}

In two spatial dimensions, while restricting ourselves to quadratic order, we can solve the problem of classification of multipole algebras. The constraint equations are
\be\la{eq:const2D}
q^i \mu_i^{I_1} =0\,,\qquad q^{i}\mu^{I_2}_{ij}=0\,,\qquad q^{ij}\mu^{I_2}_{ij}=0\,.
\ee

We consider the solutions on the case by case basis. First case is when $\mu^{I_2}_{ij}$ are non-degenerate. Then $q^i=0$ and the $\mu^{I_1}_i$ are arbitrary, which implies that the dipole moment is conserved. We have to consider a few possibilities for $q^{ij}$. To start, we parametrize the symmetric tensors using Pauli matrices and the identity matrix 
\be
\mu^{I_2} = \gamma^{I_2}_\nu \tau^\nu\,, \qquad q = \beta_\nu\tau^\nu\,,
\ee
where $\tau_\nu=(\sigma_0,\sigma_1,\sigma_3)$, $\nu=1,2,3$, and $\sigma_0$ is the identity matrix. The corresponding polynomials take form
\be
P^{I_2}_2=\gamma^{I_2}_1 (x_1^2 + x_2^2 ) + 2\gamma^{I_2}_2 x_1 x_2 + \gamma^{I_2}_3(x_1^2 - x_2^2)\,. 
\ee

The non-trivial constraints from \eqref{eq:const2D} then take form
\be\la{eq:constrainNONDEG}
\gamma^{I_2}_\nu\beta^\nu=0\,,
\ee
where $I_2=1,\ldots,k$. The number of solutions of \eqref{eq:constrainNONDEG} is $3-k$. If there are no quadratic symmetries, we find the usual symmetric tensor gauge theory, associated to the maximally symmetric multipole algebra of order $1$. If $k=1$ and $\gamma^1_{\nu}=\delta_{1,\nu}$, we find the symmetric \emph{traceless} gauge theories. These two examples are rotationally invariant with the multipole algebra of the form
\bea
&&[T_{i}, \mathcal P^{I_1}_1]=\mathcal P_0\,, \qquad [T_i, \mathcal P^{0}_2]= \delta_{iI_1}\mathcal P^{I_1}_1\,, 
\\
&& [R, \mathcal P^{I_1}_1]=\epsilon^{I_1}{}_{J_1} \mathcal P^{J_1}_1\,, \qquad [R, \mathcal P^{0}_2]=0\,,
\eea
where $R$ is the only rotation generator and $T_i$ are the translation generators. As mentioned previously, in these cases it is convenient to use $\delta_{i I_1}$ to identify the spatial indices with the multipole index $I_1$. With this identification, the corresponding gauge fields are proper tensors.

 In all other cases we find theories that break rotational symmetry since general solution for $q^{ij}_\alpha$ takes form
\be
q^{ij}_{\alpha} = \beta^\nu_{\alpha} \tau^{ij}_\nu \,,
\ee 
where $\alpha=1,2$. There are two invariant derivatives
\be
D_\alpha = q_\alpha^{ij}\p_i\p_j\,, 
\ee 
and two corresponding gauge fields $a_\alpha$. The multipole algebra takes form
\be
[T_{i}, \mathcal P^{I_1}_1]=\mathcal P_0\,, \qquad [T_i, \mathcal P^{1}_2]= f_{i,I_1}\mathcal P^{I_1}_1\,,
\ee
where $f_{i,I_1}$ are the structure constants that are determined by $\mu^{1}_{ij}$.

Next we consider $k=2$. In this case there is a single invariant derivative and a single gauge field. After gauging the Gauss law eliminates all local gauge degrees of freedom. If the quadratic symmetries are both traceless, then the only allowed $q^{ij}\propto \delta^{ij}$ and the corresponding gauge field $a_1$ is the pure trace $a_1 = \delta^{ij}a_{ij}$. This is reminiscent of the linearized dilaton coupling.

\subsubsection{Degenerate case}

More interesting structure arises in the degenerate case. The multipole symmetry contains a single linear and single quadratic term, that take form
\be\la{eq:degenerate2D}
\mu^{I_1} = (m,-n)\,, \qquad \mu^{I_2}= \left(
\begin{array}{cc}
 m^2 & -mn \\
 -mn & n^2 \\
\end{array}
\right)\,.
\ee
The corresponding polynomials are
\be
P_1^1 = m x_1 - n x_2\,, \qquad P_2^1= (m x_1 - n x_2)^2\,.
\ee
The dipole moment $q^i$ has to be orthogonal to $\mu^{I_1}$ and be a null-vector of $\mu^{I_2}$. To following vector satisfies these criteria
\be
q_1^i=(n,m)\,.
\ee
Furthermore, there are two quadrupole matrices that satisfy \eqref{eq:const2D}, which are given explicitly by
\be\la{eq:2DdipoleDEG}
q_2^{ij}=\ell_2\left(
\begin{array}{cc}
 n & \frac{m}{2} \\
 \frac{m}{2} & 0 \\
\end{array}
\right)\,, \qquad q_3^{ij} = \ell_3\left(
\begin{array}{cc}
 0 & \frac{n}{2} \\
 \frac{n}{2} & m \\
\end{array}
\right)\,,
\ee
where $\ell_2,\ell_3$ are the overall length scales, which will be determined when we discretize the theory on the lattice.

 The multipole algebra does not contain rotations and takes a simple form
\bea\la{eq:MAdipole2D1}
&&[\mathcal P^1_1, T_{\perp}]= \mathcal P_0\,,\quad [\mathcal P^{I_2}_2, T_{\perp}]= \mathcal P^1_1\,,
\\\la{eq:MAdipole2D2}
&&[\mathcal P^1_1, T_{||}]= 0\,,\quad [\mathcal P^{I_2}_2, T_{||}]= 0\,,
\eea
where $T_{||}$ and $T_{\perp}$ are translations in the direction parallel and perpendicular to $q^i$.

There are three invariant derivatives, 
\be\la{eq:invDdipole}
D_1 = q^i_1\p_i\,,\quad D_2 = q_2^{ij}\p_i\p_j\,,\quad D_3 = q_3^{ij}\p_i\p_j\,,
\ee
and three corresponding gauge fields, $a_\alpha$. There is a non-linear relation between the invariant derivatives, which takes form
\be\la{eq:relation2D}
D_1^2 = nD_2 + mD_3\,.
\ee
This relation allows to include only the terms linear in $D_1$.

The Gauss law takes form
\be
D^\dag_\alpha e_\alpha = \rho\,.
\ee
The corresponding charge configurations are illustrated in Fig.\ref{Dipole2d}. Notice, that due to the existense of the hopping operators for the $(n,m)$ dipole, the latter is fully mobile. However, the charges themselves are immobile. 

To be complete we construct a set of ``magnetic fields''. For the purpose of this work, we will call a ``magnetic field'' \emph{any} gauge-invariant combination of the gauge fields $a_\alpha$. Such construction turns out to be quite a non-trivial problem. To illustrate the sublety, we first proceed in a ``naive'' way. By inspection we find $4$ ``magnetic fields''
\bea
&& b_1 = D_1 a_2 - D_2 a_1\,, \quad  b_2 =D_1 a_3 - D_3 a_2\,,
\\
&& b_3 = D_3 a_2 - D_2 a_3\,, \quad b_4 = D_1 a_1 - n a_2 - ma_3\,.
\eea
where the first three are gauge invariant by the virtue of the commutativity of the partial derivatives, while the latter is invariant due to the non-linear relation \eqref{eq:relation2D}. It turns out, that these magnetic fields satisfy several constraints. These constraints can be obtained by applying various invariant derivatives to $b_4$. We find the following constraints
\bea
&&D_2 b_4+ D_1b_1  = mb_3
\\
&&D_3 b_4+ D_1 b_2   = -n b_3
\\
&&D_1 b_4 = 0\,.
\eea
These constraints leave a single independent component of the magnetic field. 

Finally, we could impose only a single linear polynomial symmetry and no quadratic symmetries. This will lead to a single choice of $q^i$, but no restrictions on $q^{ij}$. Such theories brake rotational symmetry.

\subsection{Anisotropic scaling}

Next, we would like to investigate the scaling properties of the two-dimensional theories. We will focus on the case when a single component of the dipole moment is conserved. The Lagrangian
\be
\mathcal L = \dot\varphi\dot\varphi + \sum_\alpha \lambda^\alpha (D_\alpha\varphi)^2
\ee
does not possess any peculiar scaling properties for generic values of $m,n$ and the coupling constants. However, there are a few interesting cases to consider. To make the scaling properties apparent we introduce the variables $\mathrm{q}=q^ix_i/|q|$ and $\mathrm{r}=\mu^1_i x^i/|\mu^1|$. In terms of these variables the Lagrangian takes form
\be\la{eq:deglag2D}
\mathcal L = \dot \varphi \dot \varphi + \lambda^\prime_1 (\p_{\mathrm{q}}\varphi)^2 + \lambda^\prime_2 (\p_{\mathrm{q}}^2\varphi)^2 + \lambda^\prime_3 (\p_{\mathrm{r}}\p_{\mathrm{q}}\varphi)^2\,,
\ee 
where $\lambda^\prime_{\alpha}$ are linear combinations of $\lambda_\alpha$, with the coefficients determined by $m,n$. The lack of $\p_{\mathrm{r}}^2\varphi$ term is a manifestation of the degeneracy of the theory. If $\lambda^\prime_2=0$ then the theory exhibits the following scaling symmetry
\be
t\rightarrow \ t\,, \,\,\, \mathrm{q}\rightarrow  \mathrm{q}\,,\,\,\, \mathrm{r}\rightarrow \beta\mathrm{r}\,,\,\,\, \varphi\rightarrow \beta^{-\frac{1}{2}}\varphi\,. 
\ee
This symmetry implies scale invariance in the direction of the conservation of the dipole moment. If such scaling symmetry is enforced, it leads to a relation between the coupling constants $\lambda_2 = - \frac{m^2}{n^2} \lambda_3$.

Next we relax the quadratic multipole symmetry. The multipole algebra, in this case, still closes and still does not contain the rotational symmetry since we have picked a preferred direction (namely, the conserved component of the dipole moment). For such theory the Lagrangian is no longer degenerate and is given by \eqref{eq:deglag2D} plus an extra term $\lambda^\prime_4(\p_{\mathrm{r}}^2\varphi)^2$. Such theory acquires an interesting scaling symmetry if we set $\lambda_2^\prime=\lambda_3^\prime=0$
\be\la{eq:lagscaleinv}
 \mathcal L_{inv} = \dot \varphi \dot \varphi + \lambda^\prime_1 (\p_{\mathrm{q}}\varphi)^2  + \lambda^\prime_4(\p_{\mathrm{r}}^2\varphi)^2\,,
\ee
where the scaling symmetry takes a highly anisotropic form
\be
t\rightarrow \beta t\,, \,\,\, \mathrm{q}\rightarrow \beta \mathrm{q}\,,\,\,\, \mathrm{r}\rightarrow \beta^{\frac{1}{2}} \mathrm{r}\,,\,\,\, \varphi\rightarrow \beta^{-\frac{1}{4}}\varphi\,. 
\ee
If such symmetry is enforced, only two charge configurations are allowed. We leave the detailed investigation of such symmetries to future work. Clearly, anisotropic scaling is only possible if the systems in question break rotational symmetry, which is often the case in condensed matter physics. It is possible that such symmetries naturally emerge close to a quantum critical point that exhibit dimensional reduction\cite{xu2005reduction, sebastian2006dimensional}.

\subsection{$U(1)$ Haah code in three dimensions} 

\begin{figure*}
\includegraphics[width=\textwidth]{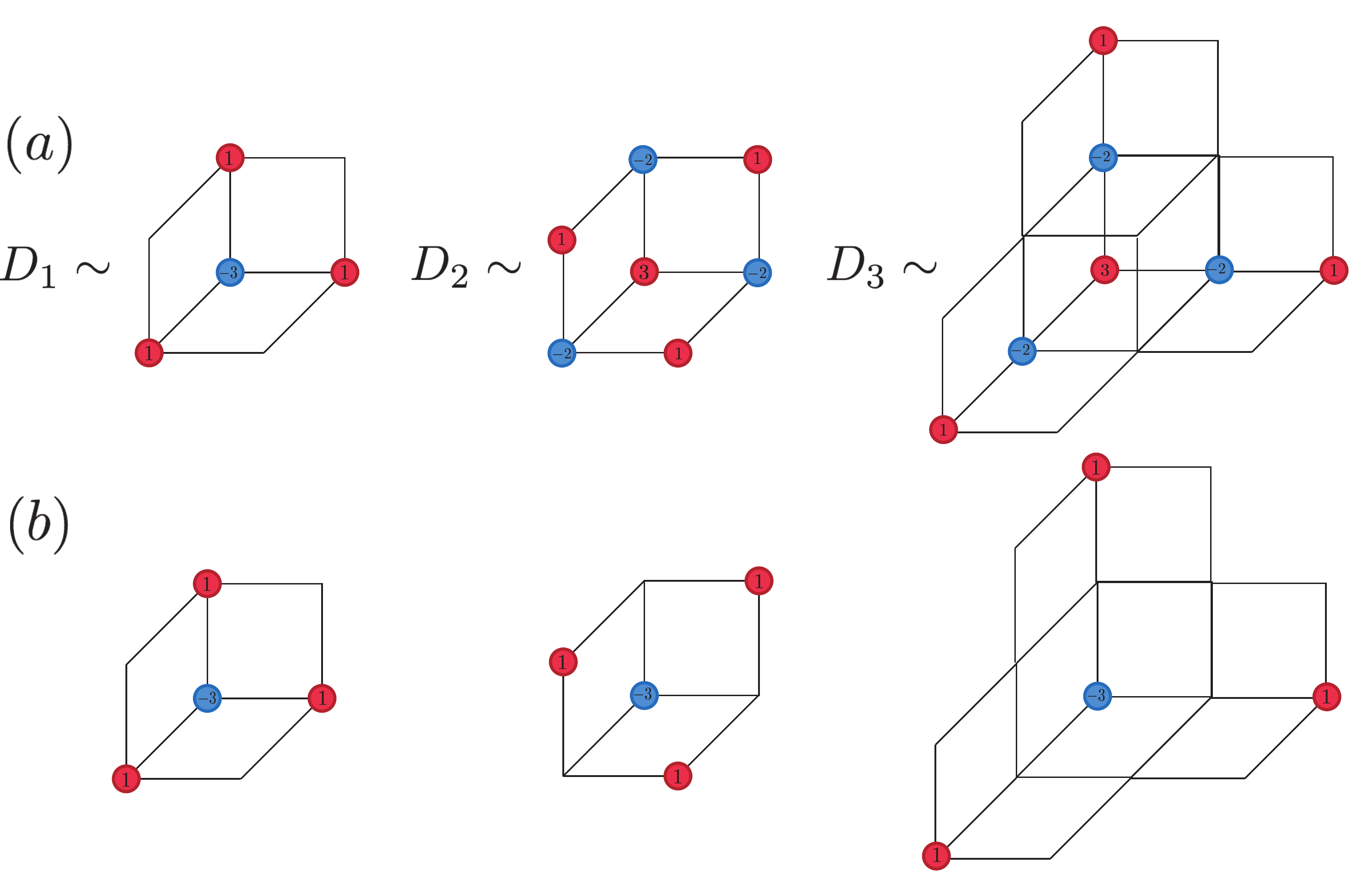}
\caption{(a) The elementary charge configurations, corresponding to $D_\alpha$, for the effective theory for the $U(1)$ Haah code \eqref{eq:HaahDerivatives}. charge configurations. (b) A different basis of elementary charge configurations. The first two configurations are precisely the ones studied in [\onlinecite{bulmash2018generalized}], while the last charge configuration is allowed by symmetries and is linearly independent from others.}
\label{Haah3D}
\end{figure*}

Next we turn to the ``$U(1)$ Haah code'' studied in [\onlinecite{bulmash2018generalized}]. We start by postulating the symmetries
\be\la{eq:Haah}
\delta \varphi = \lambda + \lambda^1_{I_1} P^{I_1}_1 + \lambda^2_{I_2} P^{I_2}_2\,, 
\ee
where
\bea
&&P^1_{1} = x_1-x_2\,,\quad P^2_{1}= x_1+x_2-2x_3\,,
\\
&&P^1_{2} = (x_1-x_2)(x_1+x_2-2x_3)\,,
\\
&& P^2_{2} =(2x_1-x_2-x_3)(x_2-x_3)\,.
\eea
The polynomials can also be represented by coefficient matrices as in \eqref{eq:mu1}-\eqref{eq:mu2}
\bea
&&\mu^{1}_i =(1,-1,0)\,, \quad \mu^2_i=(1,1-2)\,,
\\
&& \mu^{1}_{ij}= \begin{pmatrix}
 1 & 0 & -1 \\
 0 & -1 & 1 \\
 -1 & 1 & 0 
\end{pmatrix}\,,\,\,\, \mu^{2}_{ij}= \left(
\begin{array}{ccc}
 0 & 1 & -1 \\
 1 & -1 & 0 \\
 -1 & 0 & 1 \\
\end{array}
\right)\,.
\eea
The multipole algebra takes exactly the form \eqref{eq:multiexample1}-\eqref{eq:multiexample4}.
The dipole and quadrupole vectors are found by solving the constraints \eqref{eq:homoConst}-\eqref{eq:tric}. We find \emph{five} solutions which are explicitly given by
\bea
&&\bar q^i_1 = \bar \ell_0(1,1,1)\,,\qquad \bar q^{ij}_{1} = \bar \ell_1\left(
\begin{array}{ccc}
 1 & 0 & 0 \\
 0 & 0 & -\frac{1}{2} \\
 0 & -\frac{1}{2} & 0 \\
\end{array}
\right)\,,
\\
&&\bar q^{ij}_{2}=\bar \ell_2\left(
\begin{array}{ccc}
 0 & \frac{1}{2} & 0 \\
 \frac{1}{2} & 0 & 0 \\
 0 & 0 & -1 \\
\end{array}
\right)\,,\quad 
\bar q^{ij}_3 =\bar \ell_3\left(
\begin{array}{ccc}
 0 & 0 & \frac{1}{2} \\
 0 & 0 & \frac{1}{2} \\
 \frac{1}{2} & \frac{1}{2} & 1 \\
\end{array}
\right)\,,
\\
&&\bar q^{ij}_4 = \bar \ell_4\left(
\begin{array}{ccc}
 0 & 0 & 0 \\
 0 & 1 & \frac{1}{2} \\
 0 & \frac{1}{2} & 1 \\
\end{array}
\right)\,,
\eea
where $\bar \ell_\alpha$ are the overall scales, among which $\bar \ell_0$ is dimensionless, so we will set it to $1$. Recall, that these scales cannot be determined from the constraints alone. To lighten up the equations we will set all $\bar \ell_\alpha=1$, however the reader should be keenly aware that there is some freedom in overall scales. Thus we have five invariant derivatives
\be
\bar D_1= \bar q^i\p_i\,,\,\,\, \bar D_\alpha = \bar q^{ij}_{\alpha}\p_i \p_j\,.
\ee
An invariant Lagrangian of the form \eqref{eq:InvLag} can be written already at this stage. It turns out, however, that such Lagrangian is not invariant w.r.t. rotations in the $\mu^1_i-\mu^2_j$ plane. We can additionally enforce the invariance under rotations. Technically this is done by taking the linear combinations of the tensors $\bar q^{ij}_\alpha$ that are invariant under such rotations. There are two such linear combinations
\bea
&&q^{ij}_1 = \bar{q}^{ij}_1 + \bar{q}^{ij}_4=\left(
\begin{array}{ccc}
 1 & 0 & 0 \\
 0 & 1 & 0 \\
 0 & 0 & 1 \\
\end{array}
\right)\,,
\\
&&q^{ij}_2 = \bar{q}^{ij}_2 + \bar{q}^{ij}_3=\left(
\begin{array}{ccc}
 0 & \frac{1}{2} & \frac{1}{2} \\
 \frac{1}{2} & 0 & \frac{1}{2} \\
 \frac{1}{2} & \frac{1}{2} & 0 \\
\end{array}
\right)\,.
\eea
Thus we have \emph{three} invariant derivatives that take form
\be\la{eq:HaahDerivatives}
D_1 = \bar D_1\,, \quad D_2 =  q^{ij}_1 \p_i\p_j\,, \quad  D_3 =  q^{ij}_2 \p_i\p_j\,,
\ee
while the covariant derivatives are given by $ \nabla_\beta\varphi =  D_\beta\varphi +a_\beta $.

 The most general Lagrangian, consistent with the (gauged) multipole algebra takes form
\be\la{eq:HaahLag}
\mathcal L = \nabla_0 \varphi\nabla_0 \varphi -  g_{\alpha\beta}(\nabla_\alpha\varphi)(\nabla_\beta\varphi) + \ldots - \mathcal H[e,b]\,,
\ee
where $\,\ldots\,$ stands for the terms higher in derivatives, $g_{\alpha\beta}$ is the matrix of coupling constants and $\mathcal H[e,b] \propto \sum_\beta e_\beta^2 + b^2$ is the Hamiltonian for the gauge fields. 

The Gauss law takes form
\be\la{eq:GaussHaah}
\sum_\beta D^\dag_\beta e_\beta = \rho\,.
\ee
The elementary charge configurations are illustrated in Fig.\ref{Haah3D}. 

Finally, we note that there is a non-linear relation among the invariant derivatives, namely, 
\be\la{eq:relationHaah3D}
( D_1)^2 = D_2 + 2 D_3\,,
\ee
which corresponds to 
\be
q^i \otimes q^j =   q_1^{ij} + 2 q_2^{ij}\,.
\ee
These relations reduce the number of independent higher order terms and restrict the magnetic fields, but do not allow to eliminate any of the invariant derivatives.

Next we turn to the construction of the ``magnetic fields''. As before, the latter are \emph{defined} to be the gauge invariant combinations of the gauge fields. There are $4$ gauge invariant functions that we construct by inspection
\bea
&&b_4= D_1 a_1 - a_2 - 2a_3\,,\quad b_1 = D_2 a_1 - D_1 a_2\,,
\\
&&  b_2 = D_1 a_3 - D_3 a_1\,, \quad b_3 = D_3 a_2 - D_2 a_3\,.
\eea
These magnetic fields are not all independent. The magnetic field $b_4$ is gauge invariant by the virtue of the nonlinear relation between the invariant derivatives \eqref{eq:relationHaah3D}. Furthermore, it can be explicitly checked that there are three relations between the magnetic fields
\bea
&&D_2 b_4 + D_1 b_1 = 2 b_3\,,
\\
&&D_3 b_4 + D_1 b_2 = - b_3\,,
\\
&&D_1b_4=0\,,
\eea
where we have used \eqref{eq:relationHaah3D}. We again find that only one component of the magnetic field is independent.

The above theory includes the ``generalized gauge theory'' for the $U(1)$ Haah code of Ref.[\onlinecite{bulmash2018generalized}]. It appears that in Ref.[\onlinecite{bulmash2018generalized}] the authors only kept the following invariant derivatives
\be
D^{\rm BB}_1 =  D_1\,, \qquad D^{\rm BB}_2 =  D_3 - 2 D_1\,. 
\ee
We are not aware of an additional symmetry principle that would force us to discard $ D_2$. Since the Lagrangian \eqref{eq:Haah} is \emph{effective}, it must include \emph{all} terms allowed by the general principles. Addition of an extra derivative, and, therefore an extra charge configuration does not contradict the conclusion about ``fractal dynamics'' observed in [\onlinecite{bulmash2018generalized}] as we will discuss in the next Section.   

The Lagrangian \eqref{eq:HaahLag} has a hidden \emph{conformal} sliding symmetry. To see it, we introduce new variables $\mathrm{x}=\mu^{1}_i x^i/|\mu^1|$ and $\mathrm{y}=\mu^{2}_i x^i/|\mu^2|$. Then all invariant derivatives $ D_\alpha$ (and, consequently the Lagrangian) are also invariant under an infinite symmetry 
\be\la{eq:conformal}
\delta \varphi(\mathrm{z},\bar{\mathrm{z}}, x_3) = f(\mathrm{z}) + g(\bar{\mathrm{z}})\,, \quad \mathrm{z}=\mathrm{x}+i\mathrm{y}\,,
\ee
where $f(\mathrm{z})$ is holomorphic and $g(\bar{\mathrm{z}})$ is anti-holomorphic. This realization of conformal symmetry is an exotic example of a well-known ``sliding'' symmetry \cite{barci2002theory,lawler2004quantum,nussinov2005discrete}, that appears in physics of smectics \cite{o1998nonlinear} and it can be understood as a continuous version of sub-systems symmetries. This symmetry is responsible for an infinite number of conserved charges noticed in [\onlinecite{bulmash2018generalized}]. Presently, it is not clear whether sliding symmetries are a generic feature of the models invariant under the multipole algebra, or it is an accident of the Haah code.

Finally, the Lagrangian \eqref{eq:HaahLag} exhibits an anisotropic scaling symmetry, which takes form
\be
t\rightarrow \lambda t\,, \,\,\, \mathrm{x}\rightarrow \lambda^{\frac{1}{2}} \mathrm{x}\,,\,\,\, \mathrm{y}\rightarrow\lambda^{\frac{1}{2}}\mathrm{y}\,,\,\,\, x_3 \rightarrow \lambda x_3\,,\,\,\,\varphi\rightarrow \lambda^{-\frac{1}{2}}\varphi\,. 
\ee
We leave the investigation of the physical consequences of this symmetry to future work.

\subsection{Coupling to charged matter} 

We will consider charged matter, represented by a complex \emph{scalar} field. This is not the most general situation, since the matter fields will not transform under rotations, but it will serve a good illustrative purpose. The inspiration for the following construction is taken from [\onlinecite{kumar2018symmetry},\onlinecite{pretko2018gauge}].

In the previous Section we have explained how to construct the invariant derivates for an arbitrary charge conserving multipole algebra. Those derivatives were used to couple a phase field $\varphi$ minimally to the multipole gauge fields $a_\alpha$. To introduce the charged matter we view the phase field as a phase of a charged scalar, according to
\be\la{eq:phase}
\Phi = \sqrt{\rho} e^{i\varphi}.
\ee 
We will concentrate on the homogeneous multipole algebras. In this case the invariant derivatives are also homogeneous and can be ordered by the degree as follows
\be
q_\alpha^i\p_i\,,\qquad q_\alpha^{ij}\p_i\p_j\,,\qquad \ldots
\ee
The covariant derivatives of the complex scalar are defined according to
\bea\la{eq:invDcomplex1}
&&\mathcal D^1_\alpha[\Phi]=q_\alpha^i\p_i\Phi- ia_\alpha\,,
\\\la{eq:invDcomplex2}
&&\mathcal D^2_\beta[\Phi] = q_\beta^{ij}\left(\p_i\Phi \p_j\Phi - \Phi\p_i\p_j\Phi\right)-ia_\beta\,,
\\\nonumber
&&\ldots
\eea
It immediately follows that using \eqref{eq:phase} in \eqref{eq:invDcomplex1}-\eqref{eq:invDcomplex2} leads to the invariant derivatives acting on $\varphi$ that we discussed previously. The invariant Lagrangian then takes form
\be
\mathcal L = \dot \Phi^\dag\dot \Phi - g^1_{\alpha\beta} \mathcal D^1_\alpha[\Phi]^\dag D^1_\beta[\Phi] - g_{\alpha\beta}^2 \mathcal D^2_\alpha[\Phi]^\dag D^2_\beta[\Phi]-\ldots\,,
\ee
where the terms are arranged in such a way that global $U(1)$ invariance is preserved and $g_{\alpha\beta}$ is a matrix of coupling constants. It is an open problem to construct an analogue of such formalism for inhomogeneous multipole algebras as well as Lagrangians invariant up to a total derivative.



\section{Extensions} 
In this Section we discuss two extensions of the present formalism. One extension includes the point group symmetries of the lattice, whereas the other one includes charge condensation. We will also explain the relation between the present ideas and the formalism of polynomials over finite fields.

\subsection{Crystalline multipole algebra} 

 We have already observed that not all multipole algebras are consistent with continuous spatial rotations. We have also noted that the symmetry parameters have the dimension of length and, ultimately, have to be determined by the lattice constant. Keeping the lattice physics in mind, we can relax the rotational symmetry from continuous to a point group symmetry. Presently there is no general theory of the multipole algebras combined with the crystalline symmetries. Instead, we consider an example -- $C_4$ symmetry in two spatial dimensions. The polynomial symmetries compatible with $C_4$ are
\bea
&&\delta \varphi = \lambda \delta_{ij}x^ix^j + \lambda_{I_4} P_{4}^{I_4}(x)\,,
\\\nonumber
&&P_{4}^{1}=x_1^4 + x_2^4\,, \,\,\, P_{4}^{2}=x_1^3x_2 - x_1x_2^3\,,\,\,\, P^{3}_{4}=x_1^2x_2^2\,,
\eea
whereas if we were to require continuous rotational symmetry we would find a single quartic polynomial $P_4 = P_4^1 + 2P_4^3 =(x_1^2 + x_2^2)^2$.
Thus we find an interesting phenomenon: restricting continuous spatial symmetries to the crystalline ones (\emph{i.e.} reducing the symmetry), allows for increasing the multipole symmetry. This will ultimately lead to more intricate constraints on the effective Lagrangian for the ``spin-$4$'' field.


\subsection{Charge condensation} 

\begin{figure}
\includegraphics[width=3.4 in]{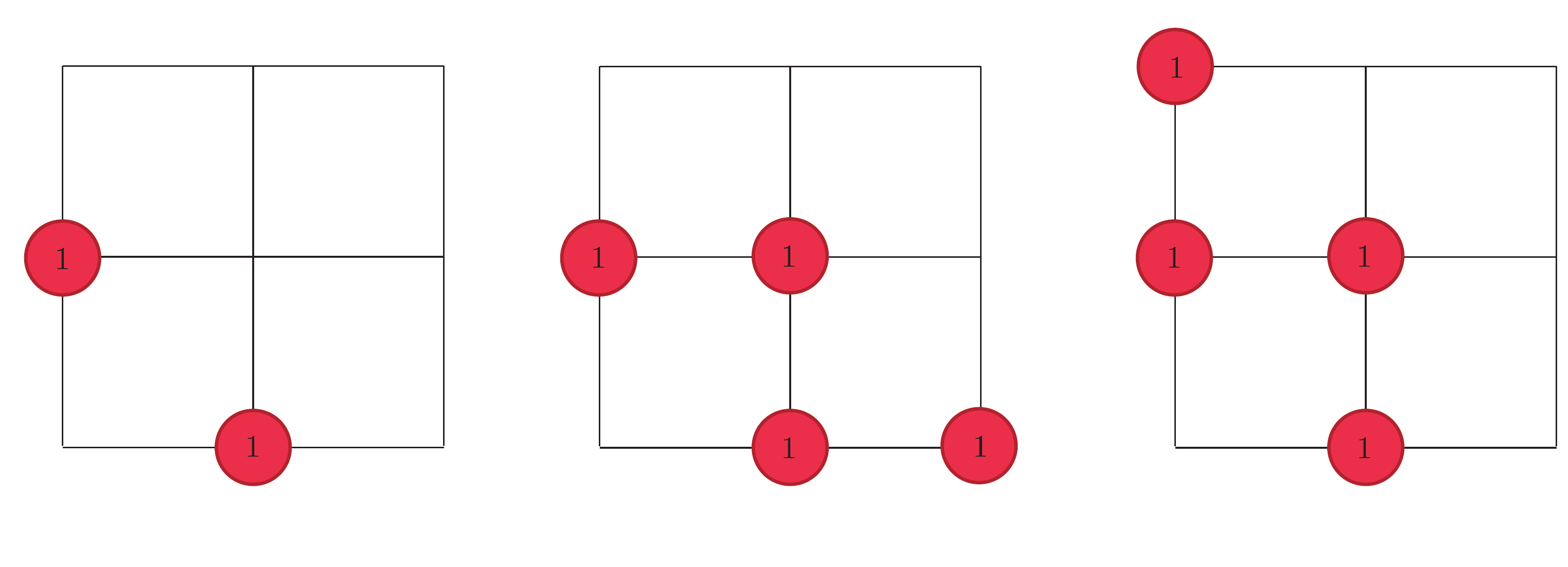}
\caption{Elementary charge configurations in the $\mathbb Z_2$ version of the theory. Since the even charges have disappeared into the vacuum the first configuration turned into a hopping operator, in the $(1,1)$ direction.}
\label{DipoleZ2}
\end{figure} 

\begin{figure}
\includegraphics[width=3.4 in]{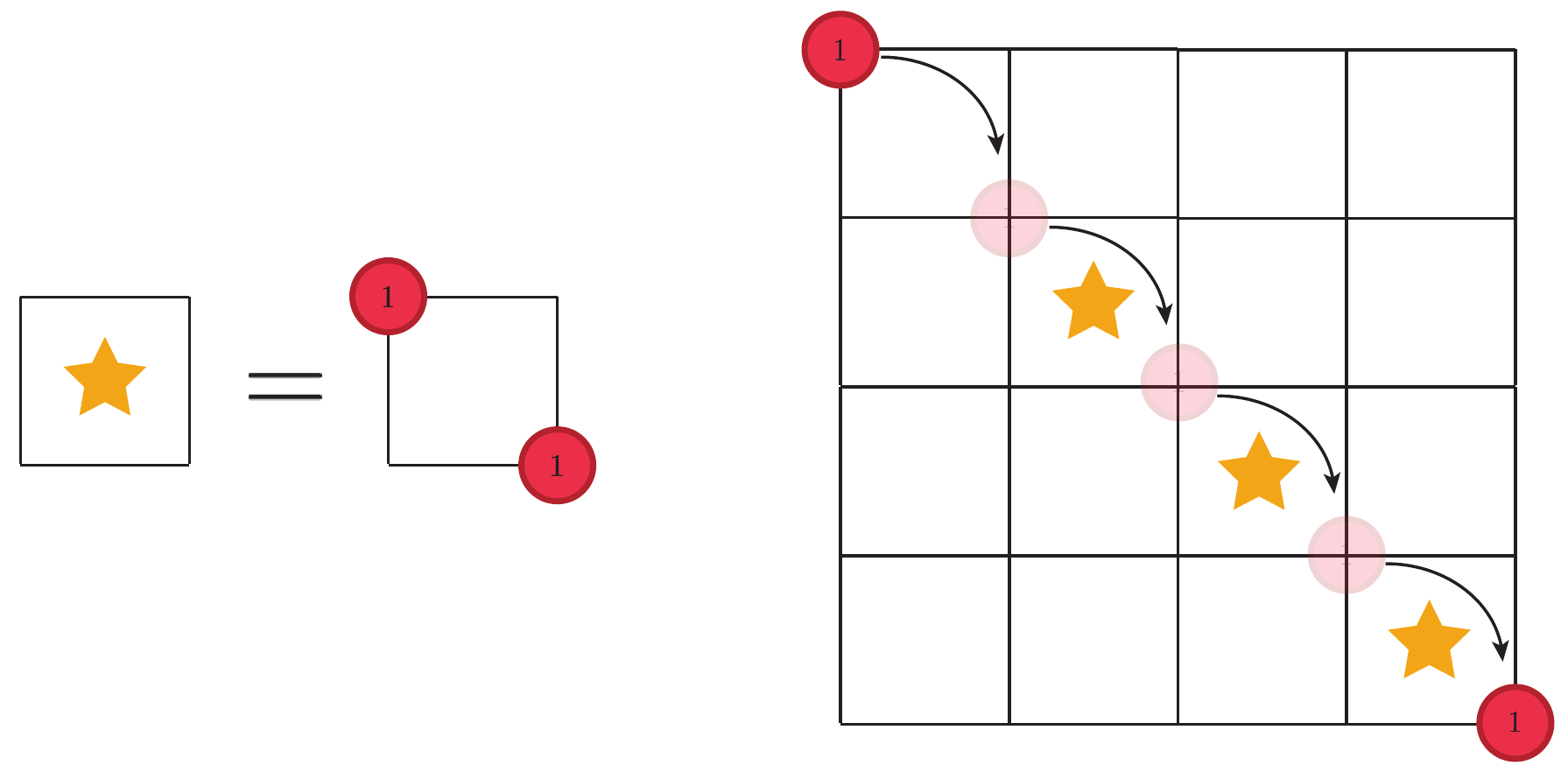}
\caption{Hopping operator, in the $(1,1)$ direction. This operator corresponds to the $D_1$ invariant derivative after the charge-2 condensation. Thus a single charge is a dimension-$1$ particle moving in the $(1,1)$ direction.}
\label{DipoleZ2hopping}
\end{figure}

Lattice fracton models usually do not have a $U(1)$ integer charge, rather they possess a $\mathbb Z_p$ symmetry, which means that the charge lattice is reduced to $\mathbb Z/p\mathbb Z$, \emph{i.e.} the excitations of charge $p$ can disappear into vacuum and are equivalent to charge $0$. This constraint is particularly effective if we have already introduced a lattice.

In light of this possibility, we will revisit the models from the previous Section. We start with a two dimensional model, characterized by the symmetry algebra \eqref{eq:degenerate2D}. In this model condensation of charge-$2$ and of charge-$3$ objects leads to dramatically different macroscopic behavior. We will consider the case when $n=m=1$. When charge-$2$ objects are condensed, we can modify Fig.\,\ref{Dipole2d}, to the $\mathbb Z_2$-valued charges, as illustrated in Fig.\,\ref{DipoleZ2}. The $(1,1)$ dipole configuration, corresponding to $D_1$, turns into an ordinary hopping operator, in the $(1,1)$ direction. The $\mathbb Z_2$ charges can be easily separated as shown in Fig.\,\ref{DipoleZ2hopping}, but only along the $(1,1)$ direction. Thus such charges are dimension-$1$ particles, capable of hopping in $(1,1)$ direction only.

\begin{figure}
\includegraphics[width=3.5 in]{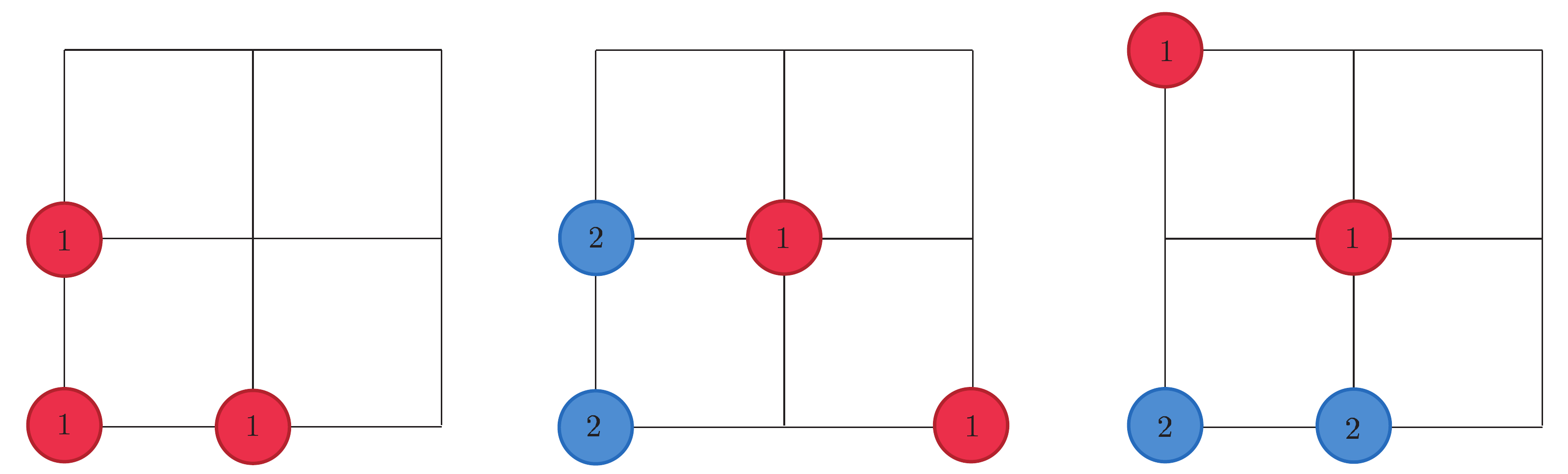}
\caption{Elementary charge configurations in the $\mathbb Z_3$ version of the degenerate $d=2$ theory. The operator, corresponding to $D_1$ is no longer a hopping operator. Note, however, that the $(1,1)$ dipole is fully mobile. This theory exhibits fractal operators.}
\label{DipoleZ3}
\end{figure}

In the $\mathbb Z_3$ case, when charge $3n$ objects are equivalent to vacuum, the $D_1$ charge configuration is no longer a hopping operator since $Q=-2\sim Q=1$. The charge configurations now take form illustrated in Fig.\,\ref{DipoleZ3}.

\begin{figure}
\includegraphics[width=3.5 in]{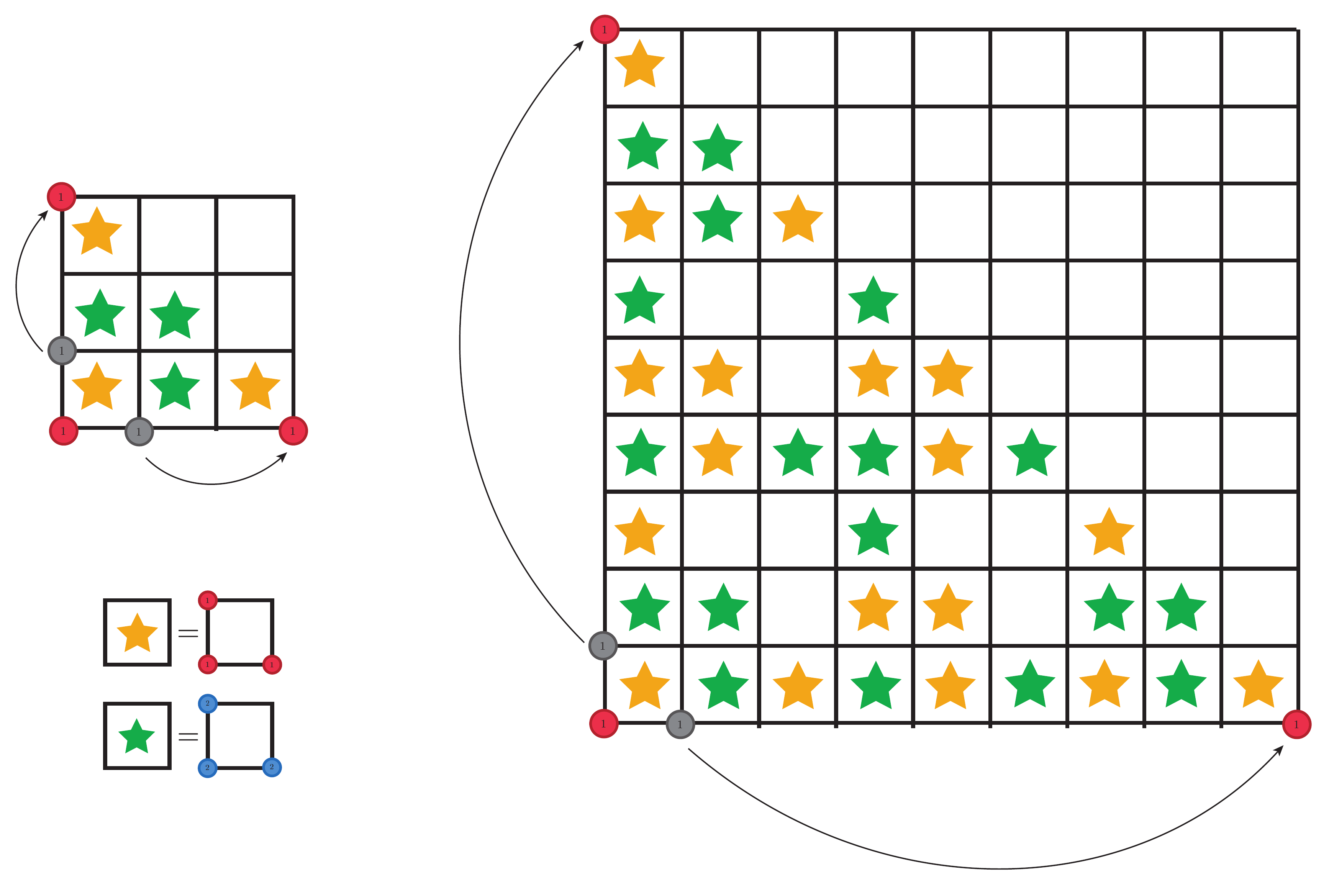}
\caption{The $(1,1)$ dipole is created in left bottom corner. The charges can be hopped by applying an operator living on a fractal structure. Yellow stars correspond to the operator $D_1$, while green stars correspond to the operator $2D_1$. It is clear that the ability to separate charges becomes sensitive to the system size since the linear size of the operators is $3^k$. The next hopping operator will be of the size $27$. Both charge $1$ and charge $2$ excitations can be hopped using similar operators. To hop charge $2$ excitations one must replace all green stars by the yellow ones and vice versa.}
\label{FractalZ3}
\end{figure}

If we try to separate the charges created by $e^{ia_1}$ we find a fractal structure (see Fig.\,\ref{FractalZ3}), which correspond to $\mathbb Z_3$ version of Sierpinski triangle. This structure has appeared in [\onlinecite{yoshida2013exotic}]. We note, however that according to the general results of [\onlinecite{haah2011local},\onlinecite{haah2013commuting}] such theories cannot be topologically ordered. Meaning, that they are not stable to perturbations that break the multipole symmetry. Rather, they should be viewed as SPTs.

\begin{figure}
\includegraphics[width=3.5 in]{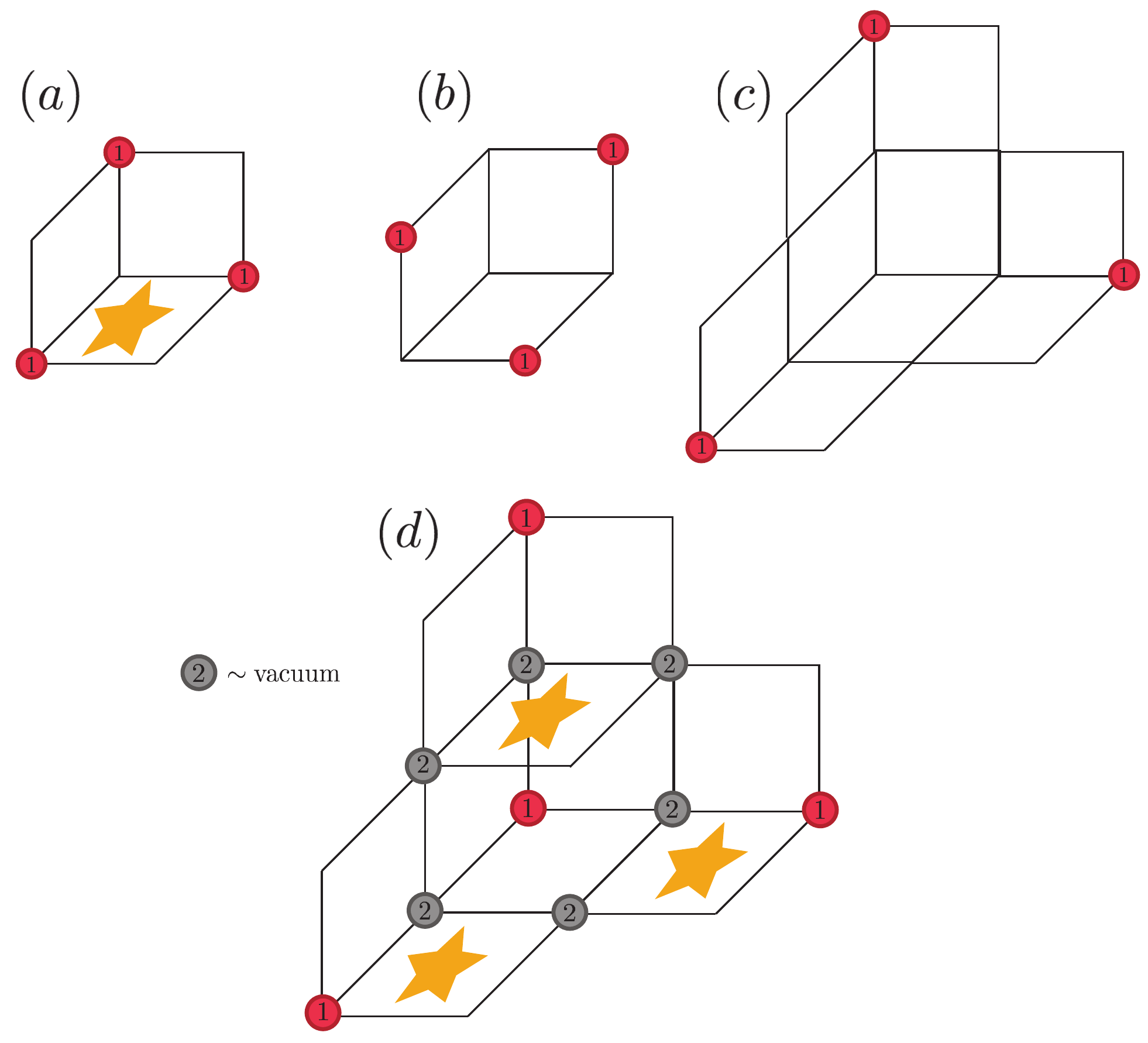}
\caption{(a)-(c). Charge configurations in the $\mathbb Z_2$ Haah code. (d). The (c) configuration can be obtained as a combination of (a) configurations, applied to plaquettes labeled by the star.}
\label{HaahZ2}
\end{figure}

Next we turn to the $U(1)$ Haah code in three dimensions. In the previous section we have found that any charge configuration is generated by a combination of $3$ basis charge configurations (see Fig.\,\ref{HaahZ2}). In the original version of the Haah's code, which is based on the $\mathbb Z_2$ charge, there are only $2$ charge configurations. The two facts are reconciled by observing that after charge-$2$ condensation one of the charge configurations can be eliminated via applying the configuration corresponding to $D_1$ multiple times as shown on Fig.\,\ref{HaahZ2}. In the $\mathbb Z_3$ version, we find three independent charge configurations.

\subsection{Polynomials over finite fields}

The original works on the fractal phases \cite{haah2011local, haah2013commuting, yoshida2013exotic} use the language of polynomials over the finite fields. In the present case, the ``creation operators'' originate in the structure of the invariant derivatives. The latter become differential operators with the coefficients in the \emph{same} finite field upon the charge condensation. In this Section we describe the relation between the formalism of polynomials over finite fields and field theoretic approach.

 To get some intuition about the possible relation we convert the graphical representation of charge configurations into polynomials as explained in\,[\onlinecite{haah2011local}, \onlinecite{yoshida2013exotic}, \onlinecite{haah2013commuting}]. In this construction one considers  formal multivariate polynomials over a finite field, say $\mathbb  Z_p$, for a prime $p$ (or over $\mathbb Z$ in the $U(1)$ case). The coefficients of the polynomials give the values of charges, while the powers of formal variables provide the coordinates. For example, $qx^n y^m$ corresponds to a charge $q$ located at position $(n,m)$.

Consider the two dimensional dipole $q^i=(1,1)$ case discussed previously. The basis charge configuration of Fig.\ref{DipoleZ3} correspond to the following polynomials (the coefficients are in $\mathbb Z_3$)
\bea\la{eq:polyZ31}
&&\mathcal H_1 = x+y + 1\,,\quad \mathcal H_2 = x^2 + xy +2y + 2\,,
\\ \la{eq:polyZ32}
&&  \quad \mathcal H_3 = y^2 + xy+ 2x +2\,.
\eea
Polynomials $\mathcal H_2$ and $\mathcal H_3$ are divisible by $\mathcal H_1$ over $\mathbb Z_3$. The coefficients in the polynomials sum up to $0$ mod $3$, which reflects the conservation of charge mod $3$. The polynomials $\mathcal H_i$ satisfy the \emph{same} relation as the invariant derivatives \eqref{eq:relation2D}
\be
\mathcal H_1^2 = \mathcal H_2 + \mathcal H_3\,.
\ee
The use of these polynomials guarantees that all charge configurations will satisfy the conservation laws. 

The hopping of the dipole can be implemented in two different ways: additively and multiplicatively. The latter is accomplished via multiplication of the polynomial by either $x$ or $y$. Indeed $x \mathcal H_1$ and $y\mathcal H_1$ correspond to the $(1,1)$ dipoles that hopped either in $x$ or in $y$ direction. The ``additive'' hopping polynomials can be constructed as $-\mathcal H_1 +x\mathcal H_1 =2 \mathcal H_1 +x\mathcal H_1 =  \mathcal H_2$. Then, indeed, $ \mathcal H_1 + \mathcal H_2 = x \mathcal H_1$. Note, that it is the additive hopping operators that correspond to the charge configurations Fig. \ref{DipoleZ3}. Since all the polynomials have a common factor, $x+y+1$, it follows that the configuration corresponding to this common factor is mobile.

Monomials $x$ or $y$ themselves are not allowed due to charge conservation. The pair creation, corresponding to $x+ 2y $ or $y+2x$ is also not allowed by the conservation of the dipole moment. The first allowed process is the triple creation, corresponding to $x+y+1$. This triple combination is mobile since the operators which hop it are also allowed by the conservation laws. The fractal operators of the Fig.\,\ref{FractalZ3} are constructed by simply considering the powers $\mathcal H_1^{3k}$. It appears that there is enough information in the field theory to construct \emph{all} of the polynomial and fractal structure after specifying the lattice and the condensation process. Alternatively, it should be possible to arrive to the same set of polynomials imposing the constraints (\emph{i.e.} multipole moment ``conservation'') directly in the polynomial language.

The Haah code is specified by a similar data. In the $\mathbb Z_2$  case the relevant polynomials are
\bea\la{eq:polyZ23D1}
&&\mathcal H_1 = x+y+z+1\,, \quad \mathcal H_2 = x^2 + y^2 + z^2 +1 \,,
\\
 && \mathcal H_3 = xy+ yz + xz +1\,. 
\eea\la{eq:polyZ23D2}
The redundancy of one of the charge configurations, in the $\mathbb Z_2$ case discussed in the previous Section, corresponds to a relation $\mathcal H_1^2 = \mathcal H_2$. One way to understand this non-linear relation is to note that in \eqref{eq:relationHaah3D} multiplication by $2$ is the same as multiplication by $0$ over $\mathbb Z_2$ (which is no longer true over $\mathbb Z_3$). This time, however, there are no mobile operators since multiplication by $x$, $y$ or $z$ takes us outside of the allowed set of polynomials.  The fractal structures are generated by $\mathcal H_1$ to various powers. 

\subsection{Multipole moments over finite fields}

General multipole moments can be constructed using the formal derivatives over finite fields. We will illustrate the construction on the example of quadratic polynomials. Consider a polynomial
\be
\mathcal H(x) = \sum_{i\leq j}h^{ij} x_i x_j + h^i x_i + h_0 \,, 
\ee
with the coefficient either in $\mathbb Z_p$ for a prime $p$ or in $\mathbb Z$. The total charge of the corresponding charge configuration is given by the sum of the coefficients within the appropriate field
\be
Q[\mathcal H]= \mathcal H\big(\{x_i\}=1\big) = \sum_{i\leq j} h^{ij} + \sum_i h^i + h^0\,.
\ee 
The conservation of charge states that we consider the polynomials satisfy $\mathcal H(\{x_i\}=1)=0$. 

The dipole moment can be evaluated as follows. Note that the power of a monomial indicates position of the charge. Thus to get a the value of the position we have to take a derivative. To this end we construct a vector of polynomials
\be
\mathcal D^k = \p^k \mathcal H = 2h^{kk} x_k +  \sum_{j\neq k}h^{kj}x_j + h^k\,.  
\ee
The dipole moment is determined by summing the coefficients in every component within the field. The sum over coefficients is formally evaluated by setting $\{x_i\}=1$
\be
d^k[\mathcal H] = \mathcal D^k \big(\{x_i\}=1\big)  = 2h^{kk} + \sum_j h^{kj} + h^k\,. 
\ee
 One has to be careful with the ``multiplication''. The symbol $2 h^{kk}$ really means $2h^{kk}\equiv h^{kk} + h^{kk}$ mod $p$. This definition automatically allows to mod out by the equivalence relations between the values of the dipole moment. We now give an example of such relations in the case of $\mathbb Z_2$. Consider a charge configuration $\mathcal H = x^2 + xy + y^2 + 1$. The total charge is $0$ over $\mathbb Z_2$ (and would be $4$ over $\mathbb Z$). The dipole polynomial takes form
\be
 \vec{\mathcal D} =
 \begin{pmatrix}
2x + y
\\
2y + x 
\end{pmatrix} = \begin{pmatrix}
 y
\\
 x 
\end{pmatrix}  \,\,\,\, \Rightarrow \,\,\,\, \vec{d}[\mathcal H] =  \begin{pmatrix}
1
\\
1
\end{pmatrix} = 
  \begin{pmatrix}
3
\\
3
\end{pmatrix}\,.
\ee
The latter equivalence between the dipole moments is a consequence of the fact that charge-$2$ excitations can be pulled out of the vacuum and shift the total dipole moment by $\delta \vec{d} = \begin{pmatrix}
2n
\\
2m
\end{pmatrix}$.

Arbitrary $k$-th multipole moment of the charge density can be defined in a similar fashion, \emph{if} we restrict ourselves to configuration with all vanishing lower moments. For example, the quadrupole moment is constructed from the matrix of second derivatives. In general, we have for the $k$-th moment $Q^{i_1i_2\ldots i_k}$
\bea
&&\mathcal Q^{i_1i_2\ldots i_k} = \p^{i_1} \p^{i_2}\ldots\p^{i_k} \mathcal H\,,
\\
&& Q^{i_1i_2\ldots i_k} =\mathcal Q^{i_1i_2\ldots i_k}\big(\{x_i\}=1\big)\,.
\eea
To check that these relations are true one has to (i) construct the $k$-th moment according to the usual definition and (ii) use the constraints that all lower moments vanish. In the case when some of the lower moments are non-zero, the construction has to be appended.

Finally, we would like to demonstrate that the multipole moments so defined are consistent with the multiplicative realization of the translations. For brevity we demonstrate this on the example of the second moment. Translation in the $l$-th direction by $n$ lattice spacings is realized via multiplication by $x^n_l$. The change in the second moment is the given by
\bea\nonumber
&&\delta_l \mathcal Q^{ij}= \p^i \p^j(x^n_l \mathcal H) - \p^i \p^j \mathcal H 
\\\nonumber
&& = \delta^i_l \p^j \mathcal H + \delta^j_l \p^i \mathcal H  + n(n-1) \delta^i_l\delta^j_l \mathcal H + (x^n_l - 1)\p^i \p^j \mathcal H\,, 
\eea
evaluating $\delta_l \mathcal Q^{ij} $ at $\{x_i\}=1$ we find
\be\la{eq:quadtranslation}
\delta_l Q^{ij} = n\delta^i_l d^j + n\delta^j_l d^i + n(n-1) \delta^i_l\delta^j_l Q = 0\,,
\ee
 This variation vanishes provided all lower moments -- total dipole and total charge in the present case -- vanish. In the language of polynomials the conservation laws are implemented as brute force constraints on the various moments of the charge density. It is not clear to us how to introduce the finite field version of the polynomial shift symmetries.

\section{Conclusions and Discussions} 

\subsection{Conclusions}

We have introduced the multipole algebra -- an extension of space(-time) symmetries that enforce conservation of certain multipole moments of the charge density. This algebra contains both spatial symmetries and, in the simplest scalar representation, the polynomial shift symmetries. We have explained how to gauge the latter in the flat space and have shown that the corresponding gauge theory satisfies a set of Gauss law constraints. These constraints imply that the local excitations correspond to certain charge configurations with prescribed moments of the charge density. In such models one encounters a difficulty in trying to separate the $U(1)$ charges away form each other. The recently studied symmetric tensor gauge theories of various kinds naturally fall into this structure and correspond to the maximally symmetric homogeneous multipole algebras. Crucially, the (gapless versions of) type-II fracton models also fall into the same category, and correspond to ``less symmetric'' multipole algebras.

We have discussed several concrete examples of the multipole algebras. The $U(1)$ version of the Haah's code fits naturally into this structure. Upon charge condensation from $U(1)$ to $\mathbb Z_2$ we find exactly the charge configurations considered in the original Haah's work. It was found that for $\mathbb Z_3$ charges there is an additional basis charge configuration that cannot be ruled out on the basis of the symmetry alone. We have also discussed a two-dimensional example where the fractal structures naturally emerge upon the charge condensation from $U(1)$ to $\mathbb Z_3$. Such $2D$ theories are gapped, but not topologically ordered. Thus, such theories should be viewed as SPTs of the multipole symmetry. Finally, an explanation relating the present construction to the formalism of polynomials over finite fields was provided.

\subsection{Discussions} 

In this final part we discuss some open problems remaining after this work. First and foremost, we were able to formulate a general structure which, upon charge condensation may (or may not) lead to the fractal operators. It is important to find the necessary and sufficient conditions, in the field theory language, for the appearance of these operators. Currently, we can only establish their presence by inspection.

As our two-dimensional example illustrates it is possible to have fractal operators without topological order. Such theories are not stable to perturbations that break the multipole algebra. At the same time we were able to reproduce the $\mathbb Z_2$ Haah code within the same framework. The latter, however, is topologically ordered and is stable against local perturbations, including the ones that break the multipole symmetry. It is not clear how to establish the existence of the topological order without going into details and comparing with the known commuting projector models.

The polynomial symmetries discussed above are clearly well-defined on an infinite plane. When the system is placed on a torus, \emph{i.e.} is subject to the periodic boundary conditions, the polynomial symmetries become inconsistent with the boundary conditions. If the field $\varphi$ is assumed to be compact, then we need to ensure that the exponents of these polynomials, $e^{iP^{I_a}_a}$, are consistent with the boundary conditions. Restricting to such polynomials will lead to the reduction in the number of symmetries. It would be interesting to see if identifying such polynomials provides the information about degeneracy on a torus as well as an indicator that signals whether the ``Higgsed'' theory is topologically ordered or not.

On a more formal field theory side, it would be interesting to develop a general procedure that allows gauging of the entire multipole algebra. Such gauging should lead to very exotic theories of gravity and/or elasticity. Partial progress on this topic has been made in regards of gauging the Bargmann algebra, which we have encountered upon studying traceless scalar charge theory\cite{andringa2011newtonian}. It will also be interesting to understand how the multipole algebra manifests itself in the theory of elasticity and its dual gauge theory along the ideas of [\onlinecite{kleinert1983double}]. We plan to address these and other questions in a forthcoming work.

\acknowledgments

It is a pleasure to thank A. Abanov, P. Ho\v{r}ava and J. Moore for stimulating discussions, and especially J. Haah, for patiently explaining his work.

A.G. was supported by Quantum Materials program at
LBNL, funded by the US Department of Energy under Contract
No. DE-AC02-05CH11231.

\bibliography{Bibliography}

\newpage

\end{document}